\documentclass[12pt,a4paper]{article}
\usepackage{jheppub}
\usepackage{natbib}
\newcommand{\nn}{\nonumber}
\newcommand{\te}{\theta^+}
\newcommand{\bt}{\overline{\theta}^+}
\newcommand{\I}{{\int\text{d}\te\text{d}\bt}}
\newcommand{\D}{\mathcal{D}}
\newcommand{\bD}{\overline{\mathcal{D}}}
\newcommand{\F}{\widetilde{\mathcal{F}}}
\newcommand{\bF}{\overline{\widetilde{\mathcal{F}}}}
\newcommand{\T}{\mathcal{T}}
\newcommand{\Tr}{\text{Tr}}
\newcommand{\BLGamma}{\mathbf{\Gamma}}
\newcommand{\BLPhi}{\mathbf{\Phi}} 
\newcommand{\BLE}{\boldsymbol{\widetilde E}}
\newcommand{\BLchi}{\boldsymbol{\chi}}
\newcommand{\BLF}{\widetilde{\boldsymbol{\mathcal{F}}}}

\title{Non-Abelian T-dualities in Two Dimensional $(0,2)$ Gauged Linear Sigma Models}
\author[a]{Nana Geraldine Cabo Bizet,}
\author[b]{Josu\'e D\'iaz-Correa}
\author[b]{and Hugo Garc\'ia-Compe\'an}
\affiliation[a]{Departamento de F\'isica, Divisi\'on de Ciencias e Ingenier\'ias,\\
Universidad de Guanajuato, Loma del Bosque 103,\\
C.P. 37150, Le\'on, Guanajuato, Mexico.}
\affiliation[b]{Departamento de F\'isica,\\
Centro de Investigaci\'on y de Estudios Avanzados del Instituto Polit\'ecnico Nacional,\\
P.O. Box 14-740, C.P. 07000, Ciudad de M\'exico, Mexico}

\emailAdd{nana@fisica.ugto.mx}
\emailAdd{josue.diaz@cinvestav.mx}
\emailAdd{hugo.compean@cinvestav.mx}
\abstract{Two dimensional gauged linear sigma models(GLSMs) with $(0,2)$  supersymmetry and  $U(1)$ gauge group possesing global symmetries are considered. For the case obtained as a reduction from the $(2,2)$ supersymmetric GLSM, we find the Abelian T-dual, comparing with previous studies. Then, the Abelian T-dual model of the pure $(0,2)$ theory is found. Instanton corrections are also discussed in both situations.  In the cases under study we explore the vacua for the scalar potential and we analyse the target space geometry of the dual model. A model with gauge symmetry $U(1)\times U(1)$ is also discussed as precursor of an interesting example with non-Abelian global symmetry. Non-Abelian T-dualization of $U(1)$ $(0,2)$ $2d$ GLSMs is implemented for models which arise as a reduction from $(2,2)$ supersymmetry; we study a concrete model with $U(1)$ gauge symmetry and $SU(2)$ global symmetry. 
It is shown that for a positive definite scalar potential, the dual vacua to $\mathbb{P}^1$ constitutes a disk. Instanton corrections to the superpotential are obtained and shown to be encoded in a shifting of the holomorphic function ${E}$. We conclude by analyzing an example with $SU(2)\times SU(2)$ global symmetry and we obtain that the space of dual vacua to $\mathbb{P}^1\times \mathbb{P}^1$ consists of two copies of the disk, also for the case of positive definite potential. Here we are able to fully integrate the equations of motion leading to non-Abelian duality, improving with respect to previous $(2,2)$ studies.
}

\keywords{T-Duality, Gauged Linear Sigma Model, $(0,2)$ Supersymmetry, non-Abelian T-duality.}

\begin{document}

\maketitle

\section{Introduction}

The study of suitable features of string theory compactifications leading to more realistic 
phenomenological scenarios has been always a great deal of interest in physics and mathematics \cite{Dine:1988yh,Candelas:1990pi,Batyrev:1994pg}. The physical aspects involve 
the reproduction of important known features of the low energy physics and the possibility to have new predictions of phenomenological interest \cite{greene1997string}. Some of these features are relevant, for instance, in the early universe or in microscopic aspects of black hole physics \cite{strominger1995les,Almheiri:2012rt}, and moreover, in the derivation of the Standard Model of particle physics and beyond \cite{Ibanez:1987sn}. The study of string compactifications with fluxes have played a central role leading to more realistic relations of string theory with phenomenological phenomena at low energies \cite{Giddings:2001yu}. A very important family of compactifications  are described by a two-dimensional non-linear sigma model (NLSM) on target spaces consisting of Calabi-Yau manifolds. These models are superconformal field theories in two-dimensions with a certain central charge with supersymmetry $(2,2)$ \cite{Gates:1984nk,Howe:1984fak}. They have many interesting features, however they lead to low energy effective field theories consistent with Grand Unification Theories in the four-dimensional spacetime with exceptional gauge groups. Much more realistic compactifications leading to $SU(5)$ or $SO(10)$ GUTs are the non-linear sigma models with $(0,2)$ supersymmetry \cite{Banks:1987cy}. This family represents a more general kind of compactifications than those of the $(2,2)$ kind (for some reviews, see \cite{Hubsch:1992nu,Distler:1995mi}). 

On the other hand, two-dimensional Gauged Linear Sigma Models (GLSMs) with $(2,2)$ supersymmetry were introduced by Witten in \cite{Witten:1993yc}, with the aim of studying solutions in string theory with the possibility to implement a change in the spacetime topology (target space) through a simple smooth variation of the parameters in the GLSM. These changes were observed in \cite{Martinec:1989in,Vafa:1988uu,Greene:1988ut,Lerche:1989uy}. More importantly these models work as a theory which interpolate among different phases which involve topology change. In some models the GLSMs lead to transitions between the Calabi-Yau (CY) phase in the infrared (IR) to the Landau-Ginzburg phase in the ultraviolet (UV) and vice-versa. In the IR the CY phase is obtained by studying the supersymmetric space of vacua of the underlying effective scalar potential. This space may have different geometries depending the specific GLSM that one is considering i.e. one may vary the number of chiral superfields and the corresponding amount of charge which carry these fields and the Abelian or non-Abelian groups of the gauge sector.  In the same reference \cite{Witten:1993yc}, there were also introduced the $(0,2)$ GLSMs in two dimensions. These models have similar properties as the $(2,2)$ models but there are many features on which they differ. For instance, the $(0,2)$ models are chiral. These models have been studied actively and very good sources can be found in  \cite{Witten:1993yc,McOrist:2010ae,Melnikov2019,Chen:2017mxp,Gu:2017nye,Alvarez-Consul:2020hbl,Chen:2017mxp}.

Different aspects of $(2,2)$ GLSMs have been worked out extensively, one of its prominent applications is the proof of Mirror Symmetry  \cite{Hori:2000kt,Hori:2003ic}. The procedure of Hori and Vafa looks to implement the Buscher-Giveon-Ro\v{c}ek-Verlinde target space (T-)duality algorithm (for some reviews, see  \cite{Rocek:1991ps,Giveon:1994fu,Quevedo:1997jb}) to $(2,2)$ GLSMs, gauging Abelian global symmetries. This procedure was successful to give a physical proof of the Mirror Symmetry correspondence. Localization properties of the partition function have been used to test Abelian T-duality in GLSMs leading to Mirror Symmetry \cite{benini1}. Moreover, the duality algorithm can be generalized to consider the gauging of a non-Abelian group in contrast to the gauging of an Abelian group.  This is termed the non-Abelian duality and many interesting traditional results have obtained in this direction \cite{delaOssa:1992vci,Giveon:1993ai,Alvarez:1994np}. More recently some interesting results involving the idea of non-Abelian duality can be found in Refs.  \cite{Sfetsos:2010uq,Lozano:2011kb,Itsios:2017cew,vanGorsel:2017goj}.

As we mentioned before $(2,2)$ GLSMs are an important tool to prove Mirror Symmetry of Calabi-Yau manifolds, in particular for the case of complete intersections of CY manifolds and toric varieties \cite{Hori:2000kt,Hori:2003ic}, and  there have been many studies in GLSMs and their applications \cite{Hori:2002zp,Hori:2006dk,Herbst:2008jq, Hori:2011pd,Hori:2013gga,Hori:2016txh,Gu:2020zpg,Gu:2019zkw,Gu:2020oeb,Gu:2020nub,Gu:2020ana,Gu:2020ivl,Chen:2020iyo}. However for GLSMs with $(0,2)$ supersymmetry the realization of Mirror Symmetry was least apparent. Certain kind of Mirror map can be defined for these models
\cite{Blumenhagen:1996gz,Blumenhagen:1996vu,Blumenhagen:1996tv,Blumenhagen:1997pp,Blumenhagen:1997vt,Blumenhagen:1997cn}. Other notions of the $(0,2)$ Mirror map are discussed in Refs. \cite{Adams:2003zy,Chen:2017mxp,Gu:2017nye,Gu:2019rty,Gu:2019byn,Alvarez-Consul:2020hbl}. In particular, in  \cite{Adams:2003zy} it was studied the Abelian GLSM with a gauged Abelian global symmetry. The authors follow the duality algorithm mentioned previously and they obtained a dual action which is also a $(0,2)$ GLSM. In particular they found the instanton contributions in the dual action which are compatible with the instanton corrections of the original $(0,2)$ GLSM. Other developments of $(0,2)$ GLSMs in different contexts can be found in Refs. \cite{Garcia-Compean:1998sla,Gadde:2013lxa,Franco:2021vxq,Franco:2022isw,Sacchi:2020pet,de-la-Cruz-Moreno:2020xop}. For a very recent overview of some important results of the GLSMs see \cite{Erick-Sharpe-2023,Franco:2022iap}.

In the context of $(2,2)$ GLSMs with $U(1)$ gauge symmetry, the possibility of gauging up a non-Abelian global symmetry was explored in \cite{CaboBizet:2017fzc,CaboBizet:2019ocy,Bizet:2021lnw}. In this article, the dual action was given and the instanton corrections of the dual action were determined. Moreover, for the Calabi-Yau phase, there were given some models in where the geometry of the target space was found. One motivation to go beyond the realm of Abelian T-duality in \cite{Hori:2000kt} comes from the fact that there is a large set of Calabi-Yau manifolds that don't constitute complete intersections but rather Grassmanians, Pfaffians or determinantal; that can be studied as Non-Abelian GLSMs \cite{Jockers:2012zr}, and a description of the symmetries in these models is of interest \cite{Hori:2011pd,Hori:2006dk,Hori:2016txh}, in particular the study of Mirror symmetry \cite{Gu:2018fpm,Gu:2019byn,Gu:2019zkw,Gu:2020oeb}.  Nevertheless one can also ask first the question, whether the T-dualities leading to mirror symmetry even for Abelian GLSMs can be generalized further, to obtain new geometric identifications. This is the question that we explore in this manuscript.

In the present article we start from the family of $(0,2)$ GLSMs considered in  \cite{Adams:2003zy}, arising from a reduction of the $(2,2)$ theory,  and we study the gauging of a non-Abelian global symmetry.   We also consider pure $(2,0)$ GLSMs models, not obtained from a supersymmetric reduction.

The present article is organized as follows: In Section \ref{sec2} a brief overview of $(0,2)$ GLSMs is given. Our aim is not to provide an extensive review but to give the notation and conventions we will follow. We also discuss the field content and interactions of these models. In Section \ref{sec3} we describe Abelian T-duality in $(0,2)$ GLSMs. We presented two examples, both of them with a single $U(1)$ gauge group and a pair of chiral superfields. In the first example we considered the case of a GLSM which is a reduction of a $(2,2)$ GLSM and in the second example we study the case of a pure $(0,2)$ GLSM. We find the equations of motion and the dual Lagrangian in both cases. Moreover, we compute the scalar potential and discuss the geometry of these vacua manifolds. The third example deals with a GLSM with $U(1) \times U(1)$ gauge group and $U(1)^4$ global symmetry group. We presented this model as a preliminary material which will be later generalized to non-Abelian T-duality in Section \ref{sec5} . In Section \ref{sec4}  we perform non-Abelian dualization for a general global symmetry group $G$. In order to be specific we particularize to $SU(2)$ global group and give its dual Lagrangian and study its vacua manifold.  Finally in Section \ref{sec5}  we consider an example with a $SU(2)\times SU(2)$ global symmetry. We conclude in Section \ref{sec6}  with our results and conclusions. At the end of the article we added Appendix \ref{appA} devoted to carry out the algorithm of T-duality at the level of superfield components. We show that for the simpler example of Abelian duality in this paper, the dual action coincides with the dual action of the reduced GLSM in the superfields language presented in Section \ref{sec3}. 
\section{Field representations of $(0,2)$ supersymmetry}
\label{sec2}

In this section, in order to be as self-contained as possible, we write an overview of the basic ingredients of two dimensional GLSMs with $(0,2)$ supersymmetry. Moreover, we will write down their corresponding Lagrangians and symmetries and we will describe the matter content and their interactions. Along the paper
we follow the notation and conventions on supersymmetric field theory as the
one in the references \cite{wessbagger,Gates:1983nr,doi:10.1142/7594}.  For reading  background material regarding $(0,2)$ GLSMs it is useful to consult the references \cite{Witten:1993yc,McOrist:2010ae,Melnikov2019}.

We start by reviewing the two-dimensional GLSM with $(0,2)$ supersymmetry and an Abelian gauge group, we will follow the original Witten's paper in GLSMs \cite{Witten:1993yc}. As usual, the coordinates of the $(0,2)$ superspace are given by $(y^0,y^1,\te,\bt)$. Where the first two are the space coordinates and the last two the fermionic counterparts.
The covariant superderivatives are given by
\begin{equation}
D_+=\partial_{\te}-i\bt\partial_+\;,\qquad\qquad\overline{D}_+=-\partial_{\bt}+i\te\partial_+\; ,
\end{equation}
where  $\partial_+:=\frac{\partial}{\partial y^0}+\frac{\partial}{\partial y^1},$ $\partial_-:=\frac{\partial}{\partial y^0}-\frac{\partial}{\partial y^1}$, $\partial_{\theta^+}
:= {\partial \over \partial \theta^+}$ and $\partial_{\overline{\theta}^+}
:= {\partial \over \partial \overline{\theta}^+}$.

The gauge covariant superderivatives $\D_+$, $\bD_+$, $\D_0$ and $\D_1$ are constructed with the following constraints:
\begin{eqnarray}
\D_0=D_0\;,&\qquad&\D_1=D_1\;,\\
\D_+=e^{-\Psi}D_+e^\Psi&=&(D_++D_+\Psi),\\
\bD_+=e^\Psi\overline D_+e^{-\Psi}&=&(\overline D_+-\overline D_+\Psi),\\
\D_0-\D_1&=&\partial_-+iV,
\end{eqnarray}
where $\Psi$ and $V$ are real functions, that constitute the gauge degrees of freedom. $V$ is the $(0,2)$ vector superfield and in the Wess-Zumino gauge they can be expanded in components as follows:
\begin{eqnarray}
V=v_--2i\te\overline{\lambda}_--2i\bt\lambda_-+2\te\bt D,\\
\label{Psiexp}
\Psi=v_+\te\bt.
\end{eqnarray}
The basic gauge invariant {field strength} $\Upsilon$ is defined as the field strength of $V$:
\begin{eqnarray}
\Upsilon&=&[\bD_+,\D_0-\D_1]V\nn \\
&=&\overline D_+(iV+\partial_-\Psi) \nn \\
\label{FS}
&=&i\overline D_+V+\partial_-\overline D_+\Psi\;.
\end{eqnarray}
In components, field strength is written as:
\begin{equation}\label{fs}
\Upsilon=-2\lambda_-+[2iD+(\partial_-v_+-\partial_+v_-)]\te+2i\partial_+\lambda_-\te\bt\;.
\end{equation}
The $U(1)$ gauge theory has a natural Lagrangian given by
\begin{equation}
L_{\tt gauge}=\frac{1}{8e^2}\I\overline\Upsilon\Upsilon,
\end{equation}
where $e$ is the gauge coupling constant.

There are two kinds of matter fields: the chiral multiplets $\Phi$ and the Fermi multiplets $\Gamma$. 
The bosonic covariant chiral fields $\widetilde\Phi$ are defined by the following constraint:
\begin{equation}
\bD_+\widetilde\Phi=0,
\end{equation}
where $\bD_+$ is the covariant derivative and consequently it has the components expansion:
\begin{equation}
\widetilde\Phi=\phi+\sqrt2\theta^+\psi_+-i\theta^+\overline{\theta}^+(\partial_++iv_+)\phi,
\end{equation}
which is defined with $\widetilde\Phi:=\Phi e^{\Psi}$, where the (uncharged) chiral superfield $\Phi$  fulfils the relation $\overline D_+\Phi=0$. The corresponding gauge invariant Lagrangian is given by
\begin{eqnarray}
L_{\tt chiral}&=&-\frac{i}{2}\I\ \widetilde\Phi^\dagger(\D_0-\D_1)\widetilde\Phi
\nn \\
&=&\overline\phi\phi D+i\overline\psi_+(\partial_-+iv_-)\psi-\sqrt2i(\lambda_-\psi_+\overline\phi-\overline\psi_+\overline\lambda_-\phi)
\nn \\
&-&\frac12[\overline\phi(\partial_-+iv_-)(\partial_++iv_+)\phi-(\partial_++iv_+)\overline\phi(\partial_-+iv_-)\phi].
\end{eqnarray}

In order to complete the rest of matter content let us introduce $\widetilde\Gamma$ which constitutes a $(0,2)$ Fermi multiplet. This multiplet satisfies the constraint:
\begin{equation}
\bD_+\widetilde\Gamma=\sqrt2\widetilde E\,,\qquad\qquad\text{then}\qquad\qquad\widetilde E=\frac{\sqrt2}2e^\Psi\overline D_+\Gamma,
\end{equation}
where $\widetilde E=E(\widetilde\Phi)$ is a holomorphic function of the  superfield $\widetilde\Phi$. Similarly, 
we can define $\widetilde\Gamma:=\Gamma e^{\Psi}$ and $\widetilde E:=Ee^\Psi$, where $\overline D_+\Gamma=\sqrt2E$. Thus, the expansion for this Fermi multiplet and the field $E$ are given by:
\begin{equation}
\widetilde{\Gamma}=\gamma-\sqrt2G\te-i\te\bt(\partial_++iv_+)\lambda-\sqrt2\widetilde{E}\bt\;.
\end{equation}
\begin{equation}
\widetilde{E}(\Phi)=E(\phi)+\sqrt2\te\frac{\partial E}{\partial \phi}\psi_+-i\te\bt(\partial_++iv_+)E(\phi).
\end{equation}

The dynamics of the Fermi field is given by the Lagrangian:
\begin{eqnarray}
L_{\tt Fermi}&=&-\frac12\I\bar{\widetilde\Gamma}\widetilde\Gamma \nn\\
&=&i\overline\gamma(\partial_++iv_+)\gamma+|G|^2-|E|^2-\left(\overline\gamma\frac{\partial E}{\partial\phi}\psi_++\frac{\partial \overline E}{\partial\overline\phi}\overline{}\psi_+\gamma\right).
\end{eqnarray}

In the case of $U(1)$ gauge theories (or in non-Abelian gauge theories with a gauge group with a $U(1)$ factor) we have an additional term in the Lagrangian given by the Fayet-Iliopoulos term
\begin{equation}
L_{D,\theta}=\frac t4\int\text d\theta^+\Upsilon|_{\bt=0}+ {\rm h.c.},
\end{equation}
where $t={\theta \over 2 \pi} +ir$, with $\theta=2\pi\Re(t)$ being an angular parameter and $r=\Im(t)$ is the Fayet-Iliopoulos parameter. 

In (0,2) theories there is a superpotential Lagrangian $L_J$, which is the $(0,2)$ analog of the superpotential term of the $(2,2)$ model. $L_J$ is of the form
\begin{equation}
L_{J}=-\frac{1}{\sqrt2}\int\text d\theta^+  \bigg(\Gamma_pJ^p(\Phi)|_{\bt=0}\bigg) - {\rm h.c.},
\end{equation} 
where $J^p=J^p(\Phi)$ is a holomorphic function of the $(0,2)$ chiral superfield $\Phi$, and $\Gamma_p$ are Fermi superfields (different from the previous ones). Moreover $J^p$ satisfies the relation $E_p(\Phi)J^p(\Phi) =0$, where of course $\tilde\D_+\Gamma_p=\sqrt2E_p$.

The scalar potential can be obtained by the usual procedure in supersymmetric theories (integrating in the superspace) and it is given by
\begin{equation}
U(\phi_i) ={e^2 \over 2}\bigg(\sum_iQ_i |\phi_i|^2 -r\bigg)^2 + \bigg(|E(\phi)|^2 + |J(\phi)|^2 \bigg),
\end{equation}
where it is clear the contributions coming the D-terms from the $(0,2)$ gauge multiplet and from the FI term. The last two terms come from the $E$ field and the last one, corresponds to the contribution from the superpotential. 

In the present article it will be considered $(0,2)$ GLSMs with a $U(1)$ gauge group, and with non-Abelian global symmetries to be gauged. Thus the dynamics of the models studied is given by the addition of all of these Lagrangians, i.e.,
\begin{equation}
L=L_{\tt gauge} + L_{\tt chiral} + L_{\tt Fermi} + L_{D,\theta}+ L_J.
\end{equation}
In the present article we will consider models without superpotential terms and therefore $L_J=0$.
\vskip 1truecm
\subsection{Reduction of the $(2,2)$ multiplets to $(0,2)$ superfields}
It is  known that certain $(0,2)$ GLSMs can be regarded as a supersymmetric reduction of $(2,2)$ GLSMs. The $(2,2)$ GLSM consists of chiral supefield $\Phi^{(2,2)}$, a vector superfield $V^{(2,2)}$ and its twisted field strength $\Sigma^{(2,2)}$. 

Next, we enumerate the steps for the decomposing $(2,2)$ multiplets terms of $(0,2)$ multiplets:

\begin{itemize}
\item{} The $(2,2)$ chiral superfield $\Phi^{(2,2)}$ can be decomposed into the $(0,2)$ chiral superfield by: $\Phi= \Phi^{(2,2)}|_{\theta^-=\overline\theta^-=0}$; and the $(0,2)$ Fermi superfield can be decomposed by: $\Gamma=\frac{1}{\sqrt2}D_-\Phi^{(2,2)}|_{\theta^-=\overline\theta^-=0}$. Both matter fields are supersymmetry reductions of the single $(2,2)$ chiral superfield.

\item{} The $(2,2)$ vector superfield $V^{(2,2)}$ gives the gauge field strength $\Upsilon=i\overline D_+(V-i\partial_-\Psi)$ by: $V-i\partial_-\Psi=-\overline D_-D_-V^{(2,2)}|_{\theta^-=\bar\theta^-=0}$. And also it gives a twisted chiral superfield $\Sigma$, which is identified as: $\bt\Sigma=-\frac{1}{\sqrt2}D_-V|_{\theta^-=\overline\theta^-=0}$; which is also simply: $\Sigma=\Sigma^{(2,2)}|_{\theta^-=\overline\theta^-=0}$.

\item{} It can be verified that if $\Phi^{(2,2)}$ has charge $Q$, then the $E$ field can be written as $E= Q \sqrt{2} \Sigma \Phi$. 

\item{} The holomorphic function $J$ can be obtained from the $(2,2)$ superpotential $W$ in the form $J = { \partial W \over \partial \widetilde\Phi}$.

\end{itemize}

\vskip 2truecm
\section{Abelian T-duality in $(0,2)$ GLSMs}
\label{sec3}
In this section we describe the T-duality for GLSMs with a $U^m(1)$ gauge group and a $U(1)^{k+s}$ global symmetry group to be gauged. We present the original model and find the T-dual model, by solving the equations of motion. For the sake of simplicity we consider the case when the superpotential $J$ of the $(0,2)$ model vanishes, thus the underlying scalar potential consist only of the $D$-term and the Fayet-Iliopoulos term. We found the dual action and the geometry of the space of dual vacua. We study two separate cases, the first one is the case in which the $(0,2)$ GLSM can be obtained by reduction from a $(2,2)$ model. The second case is the general case of a pure $(0,2)$ GLSM which cannot be obtained from a reduction. In both cases we describe their corresponding instanton corrections.  In the rest of the section we describe a particular reduced model with gauge group $U(1) \times U(1)$ and an Abelian global symmetry $U(1)^4$. This model was discussed in   \cite{Adams:2003zy} and it will be analysed in the context of non-Abelian duality in Section \ref{sec4}.
\vskip 1truecm
\subsection{GLSM with $U(1)^m$ gauge symmetry and $U(1)^{k+s}$ global symmetry}
Here we describe the Abelian T-dualization for general $(0,2)$ $U(1)^m$ GLSM with $U(1)^k$ global symmetry related to the chiral fields and $U(1)^s$ global symmetry associated to the Fermi fields. We start by writing a Lagrangian with a given number of $n$ chiral superfields $\Phi_i$ and $\widetilde{n}$ Fermi superfields $\Gamma_j$,
and a given number $m$ of $U(1)$ gauge symmetries
\begin{eqnarray}
\label{genCase}
L&=&\I\bigg\{\sum_{a=1}^m \frac{1}{8e_a^2}\overline\Upsilon_a\Upsilon_a-\sum_{i=1}^n\frac i2\overline\Phi_i e^{2 \sum_{a=1}^m Q_i^a\Psi_a}\bigg(\partial_{-}+i\sum_{a=1}^m Q_i^aV_a\bigg)\Phi_i \bigg\}\nonumber \\
&+&\I\bigg\{\sum_{i=1}^n\frac i2\overline\Phi_i\bigg(\overleftarrow{\partial}_{-}-i \sum_{a=1}^m Q_i^aV_a\bigg)e^{2\sum_{a=1}^m Q_i^a\Psi_a}\Phi_i\bigg\}  \nonumber\\
&-& \I\bigg\{\ \sum_{j=1}^{\widetilde{n}} \frac12e^{2 \sum_{a=1}^m \widetilde{Q}_j^a\Psi_a}\overline\Gamma_j\Gamma_j\bigg\} \nonumber  \\
&+&\sum_{a=1}^m \frac{t_a}{4}\int\text d\theta^+\Upsilon_a|_{\bt=0},
\end{eqnarray}
where $Q_i^a$ are the charges of the chiral superfields $\Phi_i$ and $\widetilde{Q}_j^a$ are the charges of the Fermi superfields $\Gamma_i^a$.

There are $m$ vector superfields
$V_a,\Psi_a$ with field strength $\Upsilon_a$. In principle each kinetic term
has a global phase symmetry, under which the chiral or the Fermi fields transform.
As all the superfields are distinct, one can employ the $m$ gauge symmetries to absorb $m$
of these phases, giving a total of $k+s$ global symmetries where $k=n-m$ ($n>m$) $U(1)$
global symmetries, these fields transform as follows:
\begin{eqnarray}
\delta_{\Lambda}V_a&=&-\partial_-(\Lambda_a+\overline \Lambda_a) /2,\qquad\delta \Psi=-i(\Lambda_a-\overline\Lambda_a)/2, \\
\Phi_i &&\rightarrow e^{i \sum_{a=1}^m Q_i^a \Lambda_a}\Phi_i,\qquad \Gamma_j\rightarrow e^{i \sum_{a=1}^m \widetilde{Q}_j^a \Lambda_a}\Gamma_j.
\end{eqnarray}
In general we would have the possibility to absorb with the $m$ Abelian gauge symmetries not only the global symmetries of the chiral superfields but the total amount of chiral and Fermi fields $n+ \widetilde{n}$. The master Lagrangian will remain the same with some few modifications in the sum's indices. This case will be not considered in this work.

In general one can consider a generic  number of Fermi multiplets, this is true
because the general (0,2) model, presented here, doesn't come necessarily from a SUSY reduction 
from the (2,2) theory. Therefore the Fermi multiplets are not necessarily related or
coupled to the chiral multiplets. In the opposite case when 
the chiral superfields and the Fermi superfields come both from the (2,2) reduction, the number
of  Fermi and chiral fields and their charges need to match.

Starting from (\ref{genCase}) one can construct the {\it master Lagrangian} (or also named intermediate Lagrangian) by gauging
the global symmetries and adding terms with Lagrange multipliers $\Lambda_b$ related to field strengths $\Upsilon_b$
\small
\begin{eqnarray}
L_{\tt master}&=&\I \sum_{a=1}^m \frac{1}{8e_a^2}\overline\Upsilon_a\Upsilon_a \nn \\
&-&\I\ \bigg\{\sum_{i=1}^k\frac i2\overline\Phi_i e^{2 \sum_{a=1}^{m} Q_i^a\Psi_a+2 \sum_{b=1}^{{k}} Q_{1i}^b\Psi_{1b}}\bigg(\partial_{-}+i\sum_{a=1}^m Q_i^aV_a+i\sum_{b=1}^k Q_{1i}^bV_{1b}\bigg)\Phi_i \bigg\} \nonumber \\
&+&\I\bigg\{\sum_{i=1}^k\frac i2\overline\Phi_i \bigg(\overleftarrow{\partial}_{-}-i \sum_{a=1}^m Q_i^aV_a-i \sum_{a=1}^k Q_{1i}^bV_{1b}\bigg)e^{2\sum_{a=1}^m {Q_i^a\Psi_a+ 2\sum_{b=1}^k Q_{1i}^b\Psi_{1b}}}\Phi_i\bigg\}  \nn \\
&-& \I\bigg\{\ \sum_{j=1}^{s} \frac12e^{2 \sum_{a=1}^m \widetilde{Q}_j^a\Psi_a+2 \sum_{c=1}^s \widetilde{Q}_{1j}^c\Psi_{1c}}\overline\Gamma_j\Gamma_j+ \sum_{j=s+1}^{\widetilde{n}} \frac12e^{2 \sum_{a=1}^m\widetilde{Q}_j^a\Psi_a}\overline\Gamma_j\Gamma_j\bigg\}\nonumber  \\
&+&\sum_{a=1}^m \frac{t_a}{4}\int\text d\theta^+\Upsilon_a|_{\bt=0}+\sum_{b=1}^{k}\I \Lambda_b \Upsilon_{1b}+\sum_{b=k+1}^{k+s}\I \Lambda_b \Upsilon_{1b} + {\rm h.c.} \nonumber\\
&-&\I \sum_{i=k+1}^n \bigg\{ \frac i2 \overline\Phi_i e^{2 \sum_{a=1}^n Q_i^a\Psi_a}\bigg(\partial_{-}+i\sum_{a=1}^m Q_i^aV_a\bigg)\Phi_i  \nonumber\\
&-& \frac i2  \overline\Phi_i \bigg(\overleftarrow{\partial}_{-}-i \sum_{a=1}^m Q_i^a V_a\bigg)e^{2\sum_{a=1}^m Q_i^a\Psi_a}\Phi_i\bigg\}.  
\end{eqnarray}
\normalsize
For simplicity in this expression we are assuming that the chiral superfields are not charged under the global symmetries that the Fermi superfields are charged, and vice-versa. One could choose that each of the chiral superfields to dualize it is charged only under a single
$U(1)$ global, such that $Q_{1i}^b={\delta_i}^b$, as it was done by Hori and Vafa in their
fundamental work on Mirror Symmetry as a T-duality \cite{Hori:2000kt}. There are $U(1)^{k+s}$ global symmetries, where $k+s=n-m+s$.  For models coming from supersymmetric reduction $s$ is zero and the Fermi superfield will be gauged with the same global symmetry implemented by the chiral superfields. In the general case there will be additional global symmetries arising thank to the presence of Fermi superfields to those come from the chiral superfields in the $(2,2)$ GLSM.

Let us now analyze the equations of motion from this master Lagrangian when the gauged fields are integrated. Due to the Weiss-Zumino gauge (\ref{Psiexp}), $e^{2\Psi}=1+2\Psi$. In this way, the fields $\Psi_1$, $V_1$ and $\Gamma_1$ are linear and the variation is easy performed. Carrying out the variation of the Lagrangian with respect to $\psi_{1b}$
we obtain for the field $V_{1b}$:
\begin{equation}
V_{1b}=A^{-1}_{bd}\bigg(-\frac{i}{2}\partial_-Y_-^d-R^d\bigg),
\end{equation}
where 
\begin{equation}
A_{bd}=\sum_{i=1}^k |\phi_i|^2Q_{1i}^dQ_{1i}^b, 
\end{equation}
and
\begin{equation}
R^d=\sum_{i=1}^k \left(-\frac{i}{2}\overline \Phi_i \delta_- \Phi_i Q_{1i}^d+|\Phi_i|^2\sum_{a=1}^m Q_i^a V_a Q_{1i}^d\right). 
\end{equation}
Here the new dual variable is defined by: $Y_\pm^c\equiv i\overline D_+\Lambda^c\pm iD_+\overline\Lambda^c$, and for simplicity, it has been used $\delta_-=\partial_--\overleftarrow\partial_-$.

Performing the variation of the Lagrangian with respect to $V_{1b}$, for the
component $\psi_{1b}$ we have
\begin{equation}
\psi_{1b}=A^{-1}_{bd}\bigg(-\frac{i}{2}\partial_-Y_+^d-S^d\bigg),
\end{equation}
and
\begin{equation}
S^d=\sum_{i=1}^k |\Phi_i|^2 Q_{1i}^d+2 \sum_{a=1}^m |\Phi_i|^2 Q_{1i}^d Q_{1i}^a\psi_a. 
\end{equation}

The variation with respect the component $\psi_{1d}$ for $d\in\{k+1,...,2k\}$ yields
\begin{eqnarray}
 Q_{1 j}^d\overline\Gamma_j \Gamma_j=-i \partial_- Y_-^d \rightarrow \overline\Gamma_j \Gamma_j=-Q^{-1}_{1jd}\partial_-Y_-^d.
\end{eqnarray}
These equations of motion are employed to find the dual model.
\vskip 1truecm
\subsection{A T-duality algorithm from a model coming from (2,2) reduction}
In this subsection we obtain explicit expressions for the equations of motion of a general master Lagrangian of a $(0,2)$ GLSM which is obtained by reduction of a $(2,2)$ model. In this case, there is a Fermi superfield for every chiral superfield and there could be also extra Fermi superfields. These Fermi fields have the same charges under the gauge group than the chiral superfields related to them i.e. $Q_i=\widetilde{Q}_i$ and we consider the case $s=0$, such that there are no extra Fermi fields. Then all the global symmetries will affect equally the chiral superfields and the Fermi superfields.  In this case, the duality procedure will be carry out in the fields $\Phi$ and $\Gamma$, and there are Lagrange multipliers $\overline\chi$ associated to $E$. So, the new dual field is $\F=e^{\Psi}\D_+\chi$.

\newpage 

We start from the following Lagrangian, with $n$ chiral fields and then $n$ Fermi fields (related to them) and without any extra Fermi field. This is $\widetilde n=n$:
\small
\begin{eqnarray}
&&L_{\tt master}=\I \sum_{a=1}^m \frac{1}{8e_a^2}\overline\Upsilon_a\Upsilon_a  \nn \\ 
&-& \I \bigg\{\sum_{i=1}^k\frac i2\overline\Phi_i e^{2 \sum_{a=1}^{m} Q_i^a\Psi_a+2 \sum_{b=1}^{k} Q_{1i}^b\Psi_{1b}}\bigg(\partial_{-}+i\sum_{a=1}^m Q_i^aV_a+i\sum_{b=1}^k Q_{1i}^bV_{1b}\bigg)\Phi_i  \bigg\}\nonumber \\
&+&\I\bigg\{\sum_{i=1}^k\frac i2\overline\Phi_i(\overleftarrow{\partial}_{-}-i \sum_{a=1}^m Q_i^aV_a-i \sum_{b=1}^k Q_{1i}^bV_{1b})e^{2\sum_{a=1}^m {Q_i^a\Psi_a+2\sum_{b=1}^kQ_{1i}^b\Psi_{1b}}}\Phi_i\bigg\}  \nn \\
&-& \I\bigg\{\ \sum_{j=1}^{k} \frac12e^{2 \sum_{a=1}^m Q_j^a\Psi_a+2 \sum_{b=1}^k Q_{1j}^b\Psi_{1b}}(\overline \Gamma_j+\overline \Gamma_{1j})(\Gamma_j+\Gamma_{1j})\bigg\} \nonumber  \\
&+&\sum_{a=1}^m \frac{t_a}{4}\int\text d\theta^+\Upsilon_a|_{\bt=0}+\sum_{b=1}^{k}\I \Lambda_b \Upsilon_b+\sum_{b=1}^{k}\I \bar{\chi}_b  E_b + {\rm h.c.} \nonumber\\
&-& \frac i2 \I \sum_{i=k+1}^n \bigg\{\overline\Phi_i e^{2 \sum_{a=1}^m Q_i^a\Psi_a}\bigg(\partial_{-}+i\sum_{a=1}^m Q_i^aV_a\bigg)\Phi_i  \\
&-&\overline\Phi_i \bigg(\overleftarrow{\partial}_{-}-i \sum_{a=1}^m Q_i^aV_a\bigg)e^{2\sum_{a=1}^m Q_i^a\Psi_a}\Phi_i\bigg\}+\I \sum_{j=k+1}^{\widetilde n} \frac12e^{2 \sum_{a=1}^k Q_{j}^a\Psi_{a}}{\overline \Gamma_j}\Gamma_j.   \nonumber
\end{eqnarray}
\normalsize
The terms in the last two lines are not charged under the global (gauged) symmetry so those fields behave as spectators. The main difference with the other case (without reduction) lies on the dualized fields. In this reduced model, the Fermi fields are also gauged and new Lagrange multipliers were added, these terms are located on the 5th line of previous equation.
If exists $Q\in GL(k)$ such that $(Q)_i^c:=Q_{1i}^c$, then  let be $X:=Q^{-1}$ to find the variations, which results in:
\begin{eqnarray}\label{eomU1genpsi}
\delta_{V_{1c}} S=0:\qquad \bigg(1+2\sum_{a=1}^m Q_j^a\Psi_a+2\sum_{b=1}^k Q_{1j}^b\Psi_{1b}\bigg)=-\frac{X_c^jY_+^c}{|\Phi_j|^2}\qquad \nn\\ 
\text{or }   \qquad\Psi_{1d}=\sum_{j=1}^kX_d^j\left(\frac{X_c^jY_+^c}{2|\Phi_j|^2}+\frac12+Q_j^a\Psi_a\right),
\end{eqnarray}
\begin{equation}\label{eomU1genV}
\delta_{\Psi_{1c}}S=0:\qquad -i\overline\Phi_j\Phi\delta_-\Phi_j+2\sum_{a=1}^mQ_j^aV_a|\Phi_j|^2+2\sum_{b=1}^kQ_{1j}^bV_{1b}|\Phi_j|^2=-\frac i2X_c^j\partial_-Y_-^c, 
\end{equation}
\begin{equation}\label{eomU1genG}
\delta_{\Gamma_{1j}} S=0:\qquad \bigg(1+2\sum_{a=1}^m Q_j^a\Psi_a+2\sum_{b=1}^k Q_{1j}^b\Psi_{1b}\bigg)(\overline\Gamma_j+\overline\Gamma_{1j})=-\sqrt2\F^\dagger_j.
\end{equation}
Notice that we have solved the equations for the gauged fields, we denote $\delta_fX$ to the equation of motion obtained for the field $X$.

From the derived equations of motion, one can obtain the dual Lagrangian for the $(0,2)$ models. In general it is involved to carry out this program, thus we shall consider in this work the simpler models, with some specific values of $m$, $n$ and $\widetilde{n}$. This will be developed in the following section for the case of susy  reduction and then for the pure $(2,0)$ case.

\vskip 1truecm
\subsection{GLSMs with a $U(1)$ global symmetry}
Now we consider a concrete model described by the GLSM Lagrangian with $m=1$, two chiral multiplets ($n=2$) and some Fermi superfields $\widetilde{n}$. One of each kind will act as spectator field, besides it will be considered $Q=1$.

For the implementation of the algorithm of duality we consider only the relevant terms in the Lagrangian; which we call partial Lagrangian ${\bf\Delta}L$, where the other terms as the kinetic energies of the gauge fields, the FI terms and the spectator fields are omitted. Thus the partial Lagrangian is given by:
$$
{\bf\Delta}L_{\tt original}=\I\bigg\{\frac{1}{8e^2}\overline\Upsilon\Upsilon-\frac i2\overline\Phi e^{2\Psi}\bigg(\partial_{-}+iV\bigg)\Phi+\frac i2\overline\Phi\bigg(\overleftarrow{\partial}_{-}-iV\bigg)e^{2\Psi}\Phi
$$
\begin{equation}
-\frac12e^{2\Psi}\overline\Gamma\Gamma\bigg\} +\frac t4\int\text d\theta^+\Upsilon|_{\bt=0} + {\rm h.c.}\;.
\end{equation}
From this common Lagrangian, the 2 cases: 
the reduction from $(2,2)$ and the pure $(0,2)$ case are taken.

\vskip 1truecm
\subsubsection{$(0,2)$ GLSM from a reduction of a $(2,2)$ GLSM}
As we mentioned before in the case when the $(0,2)$ model is obtained as a reduction from a $(2,2)$ model, all the chiral multiplets have associated an only Fermi field ($s=0$) and the $E$ field has a special form with the reduced fields given by 
\begin{equation}
E=iQ\sqrt2\Sigma'\Phi'\;,
\end{equation}where $\Sigma'=\Sigma|_{\theta^-=\overline\theta^-=0}$, and $\Phi'=\Phi|_{\theta^-=\overline\theta^-=0}$. Thus the gauged Lagrangian is written as
\begin{eqnarray}
\label{Lag}
{\bf\Delta}L_{\tt master}&=&\I \bigg\{-\frac i2\overline\Phi e^{2(\Psi+\Psi_1)}\bigg(\partial_-+i(V+V_1)\bigg)\Phi 
\\ 
&+&\frac i2\overline\Phi\bigg(\overleftarrow{\partial}_--i(V+V_1)\bigg)e^{2(\Psi+\Psi_1)}\Phi  \nn
\\
&-&\frac12e^{2(\Psi+\Psi_1)}(\overline\Gamma+\overline\Gamma_1)(\Gamma+\Gamma_1) +\Lambda\Upsilon_1+\overline\Upsilon_1\overline\Lambda+\overline\chi\widetilde E_1+\overline{\widetilde{E}}_1\chi\bigg\}. \nn
\end{eqnarray}
Thus, the Eqs. (\ref{eomU1genV}-\ref{eomU1genG}) substituted back into (\ref{Lag}) lead to the following Lagrangian
\begin{equation}
{\bf\Delta}L_{\tt master}=\I\bigg\{-\frac i2\frac{Y_-\partial_-Y_+}{Y_+}+\frac{|\Phi|^2\F\overline\F}{Y_+}-\bigg(\Lambda\Upsilon+\overline\Upsilon\overline\Lambda+\bar\chi E+\overline E\chi\bigg)\bigg\}.
\end{equation}
Then, after the gauge fixing $|\Phi|^2=1$ the dual Lagrangian is obtained and it can be written as:
\begin{equation}\label{dual}
{\bf\Delta}L_{\tt Dual}=\I\bigg\{-\frac i2\frac{Y_-\partial_-Y_+}{Y_+}+\frac{\F\overline\F}{Y_+}\bigg\}-\int\text d \bt\bigg(iY\Upsilon-\bar EF\bigg)+ {\rm h.c.}.
\end{equation}
This process can also be realized by components, gauging up each component field to find the scalar potential; this procedure is carried out in Appendix \ref{appA}.

Now we describe the various contributions to the scalar potential coming from the  complete dual action (\ref{dual}). This is we show the complete Lagrangian adding Fayet-Iliopoulos term, gauge action and the spectators fields. The kinetic term of the dual variable $Y$ in the first term of the dual action  $-\frac{i}{2}\frac{Y_-\partial_-Y_+}{Y_+}$, does not contribute to the scalar potential. The third term in the Lagrangian $iY\Upsilon_0$ leads to a scalar potential of the form $-2Dy_++2iv_{01}y_-+h.c.$ Moreover, the Fayet-Iliopoulos term gives rise to a potential of the form $D(\frac{it}{2}-\frac{i\overline t}{2})+v_{01}(\frac{t}{2}+\frac{\overline t}{2}).$ The gauge sector $L_{gauge}$ contributes with a term of the form $\frac{v_{01}^2}{2e^2}+\frac{D^2}{2e^2}$.We have two additional contributions from the terms $\frac{\F\bF}{Y_+}$ and $-\F E + {\rm h.c.}$ which lead to terms of the potential of the form $\frac{-2H\overline H}{y_+}$ and $-\sqrt2(HE+\overline E\overline H),$ respectively.

Thus, the scalar potential coming from (\ref{dual}) can be written as
\begin{eqnarray}
U_{\tt dual}&=&D\bigg(\frac i2(t-\overline t)-2y_++|\phi_2|^2\bigg)+\frac{D^2}{2e^2}+\frac{v_{01}^2}{2e^2}+v_{01}\bigg(\frac i2(t+\overline t)-2iy_-\bigg) \nn \\&+&\frac{H\overline H}{\Re(y)}+\sqrt2(HE+\overline E\overline H).
\end{eqnarray}
After eliminating the auxiliary fields $D$ and $v_{01}$, the potential is:
\begin{eqnarray}
U_{\tt dual}&=&\frac{e^2}{2}\bigg(-\Im(t)-\Re(y)+|\phi_2|^2\bigg)^2 \\
&+&\frac{e^2}{2}\bigg(\Re(t)+\Im(y)\bigg)^2+\frac{H\overline H}{\Re(y)}+\sqrt2(HE+\overline E\overline H)
\end{eqnarray}
which minimum condition with respect to $H$, $E$ and $\Re(y)$ gives
\begin{eqnarray}
E=0,\qquad H=0;\qquad \Re(y)=|\phi_2|^2-\Im(t).
\end{eqnarray}
Thus the minimum for the dual scalar potential is
\begin{equation}
U_{\tt dual} =0 \rightarrow|\phi_2|^2-\Re(y)=\Im(t),\label{324}
\end{equation}
while for the original theory is:
\begin{equation}
U_{\tt original} =0 \rightarrow|\phi|^2+|\phi_2|^2=\Im(t).
\end{equation}
From ec. (\ref{324}) one obtains a cone with vertex at $y_+ = - r$. Considering the $U(1)$ gauge symmetry this will lead to the line $\mathbb{R}^+$, such that the dual expected vacua is $\mathbb{R}^+\times\mathbb{R}$  while for the original model is $\mathbb{P}^1$. Notice that this dual is not the Abelian T-dual leading to the mirror pair obtained in \cite{Hori:2000kt,Adams:2003zy}. That to obtain the mirror pair we would need to add an spectator chiral superfield and gauge two $U(1)$ symmetries, one for each chiral superfield.

The superpotential of the original theory is given by \cite{Melnikov2019}:
\begin{equation}
W_{\tt original}=\frac{\Upsilon}{4\pi\sqrt2}\ln\bigg(\frac{\Sigma}{q\mu}\bigg). 
\end{equation}
We propose the following ansatz for superpotential of the dual theory:
\begin{equation}
W_{\tt dual}=iY\Upsilon-\overline EF+\beta F e^{\alpha Y},
\end{equation}
thus we have
$$
W_{\tt dual}=\frac{i\Upsilon}{\alpha}\ln\left(\frac{\overline E}{\beta}\right)
$$
\begin{equation}
=\frac{i\Upsilon}{\alpha}\ln\bigg(\frac{-i\Upsilon_0}{\alpha\beta F}\bigg).
\label{suppot}
\end{equation}
For $(0,2)$ theories coming from a reduction of a (2,2) model with  $\overline E=-iQ\sqrt2\Sigma\Phi$,  the
nonperturbative dual superpotential is written as \cite{Melnikov2019}
\begin{equation}
W_{\tt dual}=\frac{i\Upsilon}{\alpha}\ln\left(\frac{\Sigma}{\beta/(-iQ\sqrt2\Phi)}\right),
\end{equation}
where we can see by comparison with (\ref{suppot}) the choices of:
\begin{equation}
    \alpha=4i\pi\sqrt2\qquad\qquad\text{and}\qquad\qquad\beta=-iq\mu\sqrt2Q\Phi\;.
\end{equation}

\vskip1cm
\subsubsection{The case of a pure $(0,2)$ GLSM}
In this section we consider a building block $(0,2)$ model which is not coming from reduction
of a $(2,2)$ case, and study the Abelian T-duality on it. In this case we have $m=1$, $n=2$, $\widetilde{n}=1$, $k=1$ and $s=0$.

For a pure $(0,2)$ Abelian case the model is the same
\begin{equation}
{\bf\Delta}L_{\tt master}=\I\bigg\{\frac12(1+2\Psi)(-i\overline\Phi\delta_-\Phi+2V\overline\Psi\Psi-\overline\Gamma\Gamma)\bigg\}\;.
\end{equation}
But this time the field $\Gamma$ is not dualized and the gauged fields are only $V$ and $\Psi$. Thus the dual Lagrangian is given by
\begin{equation}
{\bf\Delta}L_{\tt dual}=\I\bigg\{-\frac i2\frac{Y_-\partial_-Y_+}{Y_+}+\frac{Y_+\overline\Gamma\Gamma}{2|\Phi|^2}\bigg\}+\int\text{d}\theta^+(iY\Upsilon) +\frac{t}{4}\int d\theta_+ \Upsilon + {\rm h.c.}
\label{dualpuremodel}
\end{equation}

The scalar potential is found to be the same that in the previous case discussed in Section $3.1.1$, except for the term $\frac{Y_+\overline\Gamma\Gamma}{2}$ which contributes to the scalar potential with a term of the form $y_+\overline GG-y_+\overline EE.$ Gathering all that, it results that the scalar potential of the dual theory given by
\begin{equation}
U_{\tt dual}=D\bigg(\frac i2(t-\overline t)-2y_++|\phi|^2\bigg)+\frac{D^2}{2e^2}+\frac{v_{01}^2}{2e^2}+v_{01}\bigg(\frac i2(t+\overline t)-2iy_-\bigg)+2\Re(y)(E\overline E-G\overline G).
\end{equation}
Thus, after eliminating $D$, the potential is:
\begin{equation}
U_{\tt dual}=\frac{e^2}{2}\bigg(-\Im(t)-\Re(y)+|\phi_2|^2\bigg)^2+\frac{e^2}{2}\big[\Re(t)+\Im(y)\big]^2+2\Re(y)(E\overline E-G\overline G),
\end{equation}
which minimum condition with respect to $G$, $E$ and $\Re(y)$ gives:
\begin{eqnarray}
E=0,\qquad G=0;\qquad \Re(y)=|\phi_2|^2-\Im(t).
\end{eqnarray}
Thus the minimum for the dual scalar potential is as in the previous case:
\begin{equation}
U_{\tt dual} =0 \rightarrow|\phi_2|^2-\Re(y)-\Im(t)=0.
\end{equation}
This is precisely the same equation found in the previous case (\ref{324}) and consequently the topology of the manifold of vacua is a also $\mathbb{R}^+\times\mathbb{R}$. Recall that for the original model the scalar potential reads:
\begin{equation}
U_{\tt original} =0 \rightarrow|\phi|^2+|\phi_2|^2=\Im(t),
\end{equation}
which together with the $U(1)$ gauge symmetry it constitutes a $\mathbb{P}^1$.

Thus, the ansatz for the superpotential in the dual model is
$$
W_{\tt dual}=iY\Upsilon+\beta e^{\alpha(Y+1)}
$$
\begin{equation}
=\frac{i\Upsilon}{\alpha}\bigg[\ln\bigg(\frac{-i\Upsilon}{\alpha\beta}\bigg)\bigg].
\end{equation}
In this case it was not possible to compare this to a corresponding case already discussed in the literature, but we plan to discuss is obtention elsewhere. 

\vskip 1truecm
\subsubsection{A model with two Abelian gauge symmetries}
As an example, let us apply this T-dualization procedure to the case of a $(0,2)$ GLSM coming from a reduction, as discussed in Ref. \citep{Adams:2003zy}. This is a GLSM with two gauge groups $U(1)$. We have to gauge 4 global symmetries, this is 3 chiral fields $\Phi$ and 3 Fermi fields $\Gamma$ under a $U(1)$ gauge symmetry and another 3 chiral $\widetilde\Phi$ and Fermi $\widetilde\Gamma$ with the other $U(1)$. In this case it has to be taken $m=2$, $n=6$, $\widetilde{n}=6$, $k=2$ and $s=0$. The former example has to give the same dual model, apart from the 2 spectator fields. Let us write the master Lagrangian in general:
\small
\begin{eqnarray}
L_{\tt master}=&-&\frac{i}{2}e^{2\Psi_1+2\Psi'_1}\overline{\Phi}_1\bigg(\partial_--i (V_1+V'_1)\bigg)\Phi_1+ {\rm h.c.}  \nn \\
&-& \frac{i}{2}e^{2\Psi_1+2\Psi'_2}\bar\Phi_2\bigg(\partial_--i (V_1+V_2')\bigg)\Phi_2 + {\rm h.c.}\nn \\
&-&\frac{i}{2}e^{2\Psi_2+2\Psi'_3}\overline{\widetilde{\Phi_1}}\bigg(\partial_--i (V_2+V'_3)\bigg)\widetilde{\Phi}_1 + {\rm h.c.} \nn \\
&-& \frac{i}{2}e^{2\Psi_2+2\Psi'_4}\overline{\widetilde\Phi}_2\bigg(\partial_--i (V_2+V_4')\bigg)\widetilde\Phi_2 + {\rm h.c.} \nn \\
&-&\frac12\big(\overline{\Gamma}_1+\overline{\Gamma'}_1\big)e^{2\Psi_1+2\Psi'_1}\big(\Gamma_1+\Gamma'_1\big)-\frac12\big(\overline{\Gamma}_2+
\overline{\Gamma'}_2\big)e^{2\Psi_1+2\Psi'_2}\big(\Gamma_2+\Gamma'_2\big)\nn\\
&-&\frac12\big(\overline{\widetilde\Gamma}_1+\overline{\widetilde\Gamma}'_1\big)e^{2\Psi_2+2\Psi'_3}\big(\widetilde{\Gamma_1}+\widetilde{\Gamma'_1}\big)-\frac12\big(\overline{\widetilde\Gamma}_2+\overline{\widetilde\Gamma}'_2\big)e^{2\Psi_2+2\Psi'_4}\big(\widetilde\Gamma_2+\widetilde\Gamma'_2\big)\nn\\
&+&\Lambda_1\Upsilon'_1+\Lambda_2\Upsilon'_2+\Lambda_3\Upsilon'_3+\Lambda_4\Upsilon'_4 + {\rm h.c.} + \overline\chi_1E'_1+\overline\chi_2E'_2+\overline\chi_3E'_3+\overline\chi_4E'_4+ {\rm h.c.} \nn\\
&-&\frac{i}{2}e^{2\Psi_1}\overline\Phi_3\bigg(\partial_--i V_1\bigg)\Phi_3 + {\rm h.c.} - \frac{i}{2}e^{2\Psi_2}\overline{\widetilde\Phi}_3\bigg(\partial_--i V_2\bigg)\widetilde\Phi_3 + {\rm h.c.} \nn \\ 
&-& \frac{1}{2}e^{2\Psi_1}\overline\Gamma_3\Gamma_3 -\frac{1}{2}e^{2\Psi_1}\overline{\widetilde\Gamma}_3 \widetilde\Gamma_3
+\I\bigg\{\frac{1}{8e_1^2}\overline\Upsilon_1\Upsilon_1+\frac{1}{8e_2^2}\overline\Upsilon_2\Upsilon_2\bigg\} \nn \\
&+&\frac{t_1}{4}\int\text{d}\theta^+\Upsilon_1|_{\overline\theta^+=0}+\frac{t_2}{4}\int\text{d}\theta^+\Upsilon_2|_{\overline\theta^+=0} + {\rm h.c.}
\end{eqnarray}
\normalsize
Dual fields to the Fermi multiplet are given by $F=\bar D_+ \chi$,  $\mathcal{F}=e^{\psi}F$. The scalar potential, the analysis of the supersymmetric vacua and the instanton corrections will not discussed here; since for the case of the non-Abelian global symmetry it will be discussed in detail in Section 5. 
The Lagrangian previously obtained constitutes two copies of (\ref{dual}), and that this Lagrangian is exactly the one obtained by \cite{Adams:2003zy} excluding the spectator terms.

\vskip 2truecm
\section{GLSMs with gauge group $U(1)^m$ and non-Abelian global symmetries}
\label{sec4}

In this section we construct the non-Abelian dual models of $(0,2)$ GLSMs. This can be implemented when there is a non-Abelian global symmetry present. Thus duality algorithm is realized gauging this non-Abelian symmetry and adding Lagrange multipliers which take values in the Lie Algebra of the global group. The only models considered in the present section are assumed to come from a reduction of a $(2,2)$ supersymmetric model thus the number of chiral fields $\BLPhi_i$ and the number of Fermi fields $\BLGamma_i$ coincide. And they are equally charged under the $U(1)^m$ gauge group and they are assumed to be also equally charged under the global group. Moreover in order to be as general as possible, we consider models where the total non-Abelian gauged group is $G=G_1\times \cdots \times G_S$. The Lagrangian of this model can be written as
\begin{eqnarray}
L_{\tt master}&=&\I \sum_{a=1}^m \frac{1}{8e_a^2}\overline\Upsilon_a\Upsilon_a \nn\\
&-&\I \bigg\{\sum_{I=1}^S\frac i2\BLPhi^\dagger_I e^{2 \sum_{a=1}^{m} Q_I^a\Psi_a+2 \Psi_{1I}}\bigg(\partial_{-}+i\sum_{a=1}^m Q_I^aV_a+i V_{1I}\bigg)\BLPhi_I \bigg\}  \nonumber \\
&+&\I\bigg\{\sum_{I=1}^S\frac i2\BLPhi^\dagger_I\bigg(\overleftarrow{\partial}_{-}-i \sum_{a=1}^m Q_I^aV_a-i V_{1I}\bigg)e^{2\sum_{a=1}^m {Q_I^a\Psi_a+2 \Psi_{1I}}}\BLPhi_I\bigg\}  \nn \\
&-& \I\bigg\{\ \sum_{I=1}^{S} \frac12 \big(\BLGamma^\dagger_I+\BLGamma^\dagger_{1I}\big) e^{2 \sum_{a=1}^m Q_I^a\Psi_a+2 \Psi_{1I}}\big(\BLGamma_I+\BLGamma_{1I}\big)\bigg\} \nonumber  \\
&+&\int\text d\theta^+ \sum_{a=1}^m \frac{t_a}{4}\Upsilon_a|_{\bt=0}+\I \sum_{I=1}^{S} \Tr \big(\Lambda_I \Upsilon_I\big) + {\rm h.c.}\nonumber\\
&+&\I \sum_{I=1}^{S}  \big(\BLchi^\dagger_I \BLE_I\big) + {\rm h.c.},
\end{eqnarray}
where $\BLPhi_I=(\Phi_{In_1},\dots,\Phi_{In_I})$, with $I\in\{1,\dots,S\}$, are vectors of chiral superfields, $n_I$ is the dimension of the representation of the Lie algebra of $G_I$ and
$V_{1I}=V_{1Ia}\T_a$, $\Psi_{1I}=\Psi_{1Ia}\T_a$ are superfields for each factor gauged group $G_I$, and it is assumed an inner product. In these definitions $\T^a$ are the generators of the Lie algebra of $G_I$. In the notation of the Lagrangian it is understood an inner product on the vector space indexed by the number of factors of the global group, thus we have to sum over the $S$ factors there.  
For the implementation of the duality algorithm we write the partial Lagrangian given by
\begin{eqnarray}\label{gxg}
\boldsymbol{\Delta} L_{\tt master}&=&\sum_{I=1}^S\I\bigg\{-\frac i2e_I\BLPhi^\dagger_I\delta_-\BLPhi_I+\BLPhi^\dagger\BLPhi e_IQ_I^\beta V_\beta+V_{1I}^be_IZ^b_I \nn \\
&+&\Psi_{1I}^a\bigg(-i\BLPhi^\dagger_I\T^a\delta_-\BLPhi_I+2Q_I^\beta V_\beta Z_I^a\bigg)+\Psi_{1I}^aV_{1I}^ba_I^{ab}\nn\\
&-&\frac12\big(\BLGamma_I^\dagger+\BLGamma_I^{a\dagger}\T^a\big)\big(e_I+2\Psi_{1I}^a\T^a\big)\big(\BLGamma_I+\BLGamma_I^a\T^a\big)\nn\\
&+&\bigg(V_{1I}^bY_{+a}+i\Psi_{1I}^b\partial_-Y_{-I}^a\bigg)\Tr(\T^a\T^b) \nn \\
&-&\frac{\sqrt2}{2}\bigg(\BLGamma_{1I}^{a\dagger}\T^a\T^b\BLF_I^b+\BLF^{a\dagger}_I\T^a\T^b\BLGamma_{1I}^b\bigg)\bigg\}\;,
\end{eqnarray}
which is basically the sum of Lagrangians corresponding to each factor of the global group $G_I$. In the previous Lagrangian we have the following definitions: $a^{ab}_I:=\BLPhi_I^\dagger\{\T^a,\T^b\}\BLPhi_I$, $e_I:=1_I+2\sum_{\alpha=1}^mQ_I^\alpha\Psi_\alpha$, $Z_I^a:=\BLPhi_I^\dagger\T^a\BLPhi_I$. 

The variations with respect to $V_{1I}^c$, $\Psi_{1I}^c$ and $\Gamma_{1I}^c$, give the following equations of motion:
\begin{eqnarray}
\delta_{V_{1I}^c}S=0:&&\qquad\qquad\Psi_{1I}^aa^{ca}=-Y_{+Ia}\Tr(\T^a\T^c)-e_IZ_I^c:=K_I^c, \label{eom1} \\
\delta_{\Psi_{1I}^c}S=0:&&\qquad\qquad V_{1I}^ba^{bc}_I+2Q_I^\beta V_\beta Z_I^c-i\Phi_I^\dagger\T^c\delta_-\Phi_I+i\partial_-Y_{-aI}\Tr(\T^a\T^c) 
\nn \\
&&\qquad\qquad-(\Gamma^\dagger_I+\Gamma_I^{a\dagger}\T^a)\T^b(\Gamma_I+\Gamma_I^c\T^c)=0, \label{eom2}\\
\delta_{\Gamma_{1I}^c}S=0:&&\qquad\qquad-\frac12(\Gamma_I^\dagger+\Gamma_{1I}^{a\dagger}\T^a)(e_I+2\Psi_{1I}^b\T^b)\T^c-\frac{\sqrt2}{2}\F_I^{a\dagger}\T^a\T^c=0.\label{eom3}
\end{eqnarray}

Thus the corresponding partial dual Lagrangian becomes
\begin{eqnarray}
\boldsymbol{\Delta} L_{\tt dual}&=&\sum_{I=1}^S\I\bigg\{-\frac i2e_I\BLPhi_I^\dagger\delta_-\BLPhi_I+\BLPhi^\dagger_I\BLPhi_Ie_IQ_I^\beta V_\beta +\BLF_I^\dagger X_I^{-1}\BLF_I \nn \\
&+&\frac{\sqrt 2}{2}(\BLF_I^\dagger\BLGamma_I+\BLGamma_I^\dagger \BLF_I) \bigg(-i\BLPhi_I^\dagger\delta_-\T^a\BLPhi_I + 2Q_I^\beta V_\beta Z_I^a\bigg) \nn \\
&\times& \bigg(-Y_{+Ib}\Tr(\T^b\T^c)-e_IZ_I^c\bigg)b_{ac}\bigg\},
\end{eqnarray}
where $b^{ac}$ is the inverse of $a^{cd}$, and $X_I:=e_I+2\T^aK_I^a=e_I-2\T^ae_IZ_I^cb^{ca}-2\T^aY_{+b}\Tr(\T^b\T^c)b^{ca}$. Still it is necessary to remove the original chiral fields $\BLPhi$ from the Lagrangian, step which will be implemented through the process of gauge fixing. Up to this point we have used generic well behaved Lie groups $G_i$, it hasn't been necessary to  specify the groups.

\vskip 1truecm
\subsection{A model with $SU(2)$ global symmetry}
\label{41}
Let us consider the case of a model with global symmetry $G=SU(2)$. Before proceeding, some definitions and conventions related to the $SU(2)$ group are introduced for future reference. Hence, the following relations hold: $\Tr(T^aT^b)=2\delta^{ab}$, $\{T^a,T^b\}=2\delta^{ab}\mathbf{I_d}$ for the generators of the Lie algeba of the group $SU(2)$; in this way we have the relations $a^{ab}=2|\Phi|^2$ and $b^{bc}=\frac{1}{2|\Phi|^2}$. Consequently, with $e_I:=1_I+2\sum_{\alpha=1}^mQ_I^\alpha\Psi_\alpha$, it is obtained that $X_I:=e_I\mathbf{I_d}-T^a\frac{e_IZ^a_I+2Y_+^a}{|\Phi_I|^2}$.

For the specific model with a $U(1)$ gauge group and an $SU(2)$ global symmetry the partial Lagrangian is given by
\begin{eqnarray}
\boldsymbol{\Delta} L_{\tt master}&=&\sum_{I=1}^s\I\bigg\{\bigg(-i\BLPhi_I^\dagger\delta_-\T^a\BLPhi_I+Q_\beta V^\beta\bigg)\nn \\
&\times&\bigg(e_I|\BLPhi_I|^2-\frac{e_I}{|\BLPhi_I|^2}Z^aZ_a-Y_+^a\frac{Z^a}{|\BLPhi_I|^2}\bigg) \nn\\
&+&\BLF^\dagger X^{-1}\BLF+\frac{\sqrt2}{2}(\BLF_I^\dagger\BLGamma_I+\BLGamma_I^\dagger\BLF_I)-\frac i2e_I\BLPhi_I^\dagger\delta_-\BLPhi_I\bigg\}.
\end{eqnarray}
For future convenience the fields $\BLPhi=\begin{pmatrix}\Phi_1 \\ \Phi_2\end{pmatrix}$, which are 2 complex fields, can be redefined in terms of new fields $Z_0,Z_1,Z_2,Z_3$:
\begin{equation}
Z_0=\overline\Phi_1\Phi_1+\overline\Phi_2\Phi_2,\qquad Z_1=2\Re(\overline\Phi_1\Phi_2),\qquad Z_2=2\Im(\overline\Phi_1\Phi_2),\qquad Z_3=\overline\Phi_1\Phi_1-\overline\Phi_2\Phi_2,
\end{equation}

Then, the original chiral fields can be eliminated by gauge fixing the $Z$'s, these are 4 real constants and the products are written as sums of chiral fields
\begin{equation}\label{phis}
\overline\Phi_1\Phi_1=\frac{Z_0+Z_3}{2},\qquad \overline\Phi_1\Phi_2=\frac{Z_1+iZ_2}{2},\qquad \overline\Phi_2\Phi_2=\frac{Z_0-Z_3}{2}.
\end{equation}
Thus, with the partial gauge fixing: $\Phi_I^\dagger T^b\partial_-\Phi_I=\partial_-\Phi_I^\dagger T^b\Phi_I$, for $b\in\{0,1,2,3\}$, the partial dual Lagrangian has the following form
\begin{eqnarray}
\boldsymbol{\Delta} L_{\tt dual}&=&\sum_{I=1}^s\I\bigg\{Q_\beta V^\beta\bigg(e_I-e_I\frac{Z^aZ_a}{Z_0}-\frac{Y_+^aZ^a}{Z_0}\bigg) \nn \\
&+&\BLF^\dagger \Big(e_I\mathbf{I_d}-\frac{\T^a}{Z_0}(e_IZ^a_I+2Y_+^a)\Big)^{-1}\BLF \nn
\\
&+&\frac{\sqrt2}{2}\bigg(\BLF_I^\dagger\BLGamma_I+\BLGamma_I^\dagger\BLF_I\bigg)\bigg\}+\frac{t}{4}\int\text d\theta^+\Upsilon|_{\bt=0}. 
\end{eqnarray}
To write the scalar potential in a more convenient form, it can be defined the variable $u^a=2\frac{y_+^a}{Z_0}+\frac{Z_a}{Z_0}$. Therefore, the new dual coordinate is denoted as $u^a$. As a result, considering all the corresponding terms, the scalar potential can be expressed as follows
\begin{eqnarray}
U_{\tt dual}&=&\frac{-2}{1-u^au_a-2\frac{Z^aZ_a}{Z_0^2}+2\frac{Z_au^a}{Z_0}} \\
&\times&\bigg[\overline H_1H_1+\overline H_2H_2+\bigg(\overline H_1H_1-\overline H_2H_2\bigg)u^3+\overline H_2H_1\overline u^{12}+\text{h.c.}\bigg]\nn\\
&+&\sqrt2\bigg[\overline H_1E_1(\phi)+\overline H_2E_2(\phi)+H_1\overline E_1(\phi)+H_2\overline E_2(\phi)\bigg] -\frac{iQv_-\partial_+u_-^aZ_a}{Z_0^2}\nn\\
&+&2Q^2v_-v_+\bigg(1-\frac{Z^aZ_a}{Z_0^2}\bigg)+2QD\bigg(1-\frac{Z^aZ_a}{2Z_0^2}-\frac{u^aZ_a}{2Z_0^2}\bigg)+ \frac{D^2}{2e}-rD. \nn
\end{eqnarray}
\normalsize
After eliminating the auxiliary field $D$ we have for the potential the following form
\begin{eqnarray}
U_{\tt dual}&=&\frac{-2}{1-u^au_a-2\frac{Z^aZ_a}{Z_0^2}+2\frac{Z_au^a}{Z_0}}\bigg[\overline H_1H_1+\overline H_2H_2+\bigg(\overline H_1H_1-\overline H_2H_2\bigg)u^3 \nn\\
&+&\overline H_2H_1\overline u^{12}+\text{h.c.}\bigg] +\sqrt2\bigg[\bar H_1E_1(\phi)+\bar H_2E_2(\phi)+H_1\bar E_1(\phi)+H_2\bar E_2(\phi)\bigg] \nn \\
&-&\frac{iQv_-\partial_+u_-^aZ_a}{Z_0^2}+2Q^2v_-v_+\bigg(1-\frac{Z^aZ_a}{Z_0^2}\bigg) \nn \\
&-&\frac e2\bigg(1-\frac{Z^aZ_a}{2Z_0^2}-\frac{u^aZ_a}{2Z_0^2}+r\bigg)^2.
\end{eqnarray}
In order to find the minimum through derivatives of $H$'s it is found the vacuum condition $U=0$,
and integrating the fields $v_-$ and $v_+$ we have
\begin{eqnarray}
&0&=\frac e2\bigg(r+1+\frac{Z^aZ_a}{2Z_0^2}-\frac{u^aZ_a}{2Z_0^2}\bigg)^2 \nn \\
&+&\frac{u^cu_c-\frac{2u^cZ_c}{Z_0}+\frac{2Z^aZ_a}{Z_0^2}-1}{1-u^cu_c} \nn \\
&\times& \bigg[|E_1|^2(1-u_3)+|E_2|^2(1+u_3)-E_1\overline E_2\overline u_{12}-E_2\overline E_1u_{12}\bigg]\;.
\end{eqnarray}
If $A=\frac{-2y_+^by_+^b}{1-u_cu^c}\begin{pmatrix}u_3-1&u_{12}\\\bar u_{12}&-1-u_3\end{pmatrix}$, then this previous condition is rewritten in the following form
\begin{equation}
U_{\tt dual}= \frac e2\bigg(\Im(t)-y_+^aZ_a\bigg)^2+(\overline E_1\quad\overline E_2)A\begin{pmatrix}E_1\\E_2\end{pmatrix}=0\;.
\label{potentiallast}
\end{equation}
This analysis is valid in the IR when the vector fields are integrated.

For future convenience let us write the original scalar potential:
\begin{equation}
U_{\tt original}=\frac{e^2}{2}\left(\sum_iQ_i|\phi_i|^2-r\right)^2+\sum_a|E_a|^2.
\end{equation}
The vacua manifold is characterized by the 3 coordinates: $y_+^a$, and there's one equation for the vacua, thus it is a two dimensional surface. The $Y_-$ term doesn't appear on the Lagrangian, then $y_-$ isn't a coordinate in the potential. The eigenvalues of the matrix $A$ are: $\lambda_\pm=\frac{2y_+^ay_+^a}{1\mp\sqrt{u^au^a}}$. So due to $A$ be Hermitian, there exists a unitary matrix $P$ such that $A=P^\dagger DP$, and $D$ is the diagonal matrix with eigenvalues as entries; therefore:
\begin{equation}
\BLE^\dagger A\BLE=\BLE^\dagger P^\dagger DP\BLE=(P\BLE)^\dagger D(P\BLE)=\lambda_+|(P\BLE)_+|^2+\lambda_-|(P\BLE)_-|^2,
\end{equation}  
which is a quadratic form. Thus, the vacua is made up of 3 regions\footnote{The point $y^a=0$ is also solution only if $\Im(t)=0$.} depending on whether $y_+^ay_+^a$ is greater than, less than or equal to $Z^ay_+^a$, these regions are: the inside of a sphere, the outside of it and the sphere shell. Thus we have three cases:
\begin{itemize}
\item{} Region 1:
\begin{equation}
 y^ay^a<Z^ay^a,\qquad \quad\Im(t)=y_+^aZ^a\quad\text{and}\quad |\BLE|^2=0.
\end{equation}
\item{} Region 2:
\begin{equation}
y^ay^a=Z^ay^a,\qquad \quad\Im(t)=y_+^aZ^a\quad\text{and}\quad |(P\BLE)_-|^2=0.
\end{equation}
\item{} Region 3:
\begin{equation}
y^ay^a>Z^ay^a,\qquad  \quad\frac e2(\Im(t)-y_+^aZ_a)^2+\lambda_-|(P\BLE)_-|^2=-\lambda_+|(P\BLE)_+|^2. 
\end{equation}
\end{itemize}
This can be also written $\Im(t)=y_+^aZ^a-\sqrt{\frac{-2}{e}\BLE^\dagger A\BLE}$,
where
\begin{equation}
(P\BLE)_\pm=\frac{\mp u_{12}E_1+(\sqrt{u_cu^c}\pm u_3)E_2}{\sqrt{2\sqrt{u_cu^c}(\sqrt{u_cu^c}\pm u_3)}},
\end{equation}
and consequently
\begin{eqnarray}
&&|(P\BLE)_\pm|^2=\\
&=&\frac{|u_{12}|^2|E_1|^2+(\sqrt{u_cu^c}\pm u_3)^2|E_2|^2\mp2(\sqrt{u_cu^c}\pm u_3)[u_1\Re(E_1\overline E_2)-u_2\Im(E_1\overline E_2)]}{2\sqrt{u_cu^c}(\sqrt{u_cu^c}\pm u_3)}. \nn
\end{eqnarray}
For the regions 1 and 2, the potential is semi-definite positive and the surface $y_+^a Z_a=\Im(t)$ is a plane inside the sphere, thus this is a disk $\bf D$, where the modulus $r=\Im(t)$ determines the size of the disk  (as its relative position inside the sphere). While for the case outside of the sphere, one has a more sophisticated surface, whose boundary is the same as the one of the disk, with the topology $\mathbb{R}^2 \diagdown{\bf D}$.

Until this point the superfield $E$ has been arbitrary and it is a function only on $\Phi$ (which is gauge fixed) and other parameters; however, it can be chosen a particular form of it that comes from the $(2,2)$ reduction:
\begin{equation}
\left(\begin{array}{c}E_1 \\ E_2\end{array} \right)=\Sigma\left(\begin{array}{c}\Phi_1 \\ \Phi_2\end{array}\right)\;,
\end{equation}
where $\Sigma=\sigma+\sqrt2\te\overline\lambda_+-i\te\bt\partial_+\sigma$, this means for region 1: $|\sigma|^2=0$. 

\vskip 1truecm

It is remarkable that in this case the equations of motion (\ref{eom1}),(\ref{eom2}) and (\ref{eom3}) can be exactly solved, without requiring to project out to an Abelian component (or to particularize to a semichiral vector field) as in the $(2,2)$ supersymmetric case \cite{CaboBizet:2017fzc}.

\subsubsection{Instanton correction}
The Lagrangian with the instanton correction is given by
\begin{eqnarray}
L_{\tt dual}&=&\sum_{I=1}^s\I\bigg\{Q_\beta V^\beta\bigg(e_I-e_I\frac{Z^aZ_a}{Z_0}-\frac{Y_+^aZ^a}{Z_0}\bigg)  \nn \\
&+&\BLF^\dagger \Big(e_I\mathbf{I_d}
-\frac{\T^a}{Z_0}(e_IZ^a_I+2Y_+^a)\Big)^{-1}\BLF+\frac{\sqrt2}{2}\bigg(\BLF_I^\dagger\BLGamma_I +\BLGamma_I^\dagger\BLF_I\bigg)\bigg\} \nn \\
&+&\int\text d\theta^+\bigg\{\frac t4\Upsilon|_{\bt=0}+\BLF^\dagger\boldsymbol{\beta}e^{\alpha^bY_b}\bigg\}, 
\end{eqnarray}
where the last term is the instanton correction and its contribution to the bosonic scalar potential is:
\begin{equation}
\int\text{d}\theta^+ \BLF^\dagger\boldsymbol{\beta}e^{\alpha^bY_b}=-\sqrt2\big(\overline H_0\beta^0+\overline H_1\beta^1\big)e^{\alpha_by_+^b}.
\end{equation}
Then, the new vacua equation is:
\begin{equation}
\frac e2\bigg(\Im(t)-y_+^aZ_a\bigg)^2+\bigg(\BLE-e^{\alpha_by_+^b}\boldsymbol\beta\bigg)^\dagger A\bigg(\BLE-e^{\alpha_by_+^b}\boldsymbol\beta\bigg)=0\;,
\end{equation}
and similarly, when $0>y_+^ay_+^a+y_+^aZ_a$ the solution gives:
\begin{equation}
\frac e2\bigg(\Im(t)-y_+^aZ_a\bigg)^2=0\qquad\text{and}\qquad |\varepsilon_1|^2(1-u_3)+|\varepsilon_2|^2(1+u_3)-\varepsilon_1\overline \varepsilon_2\overline u_{12}-\varepsilon_2\overline \varepsilon_1u_{12}=0,
\end{equation}
where $\boldsymbol\varepsilon=\BLE-e^{\alpha_by_+^b}\boldsymbol\beta$.
Notice that the effect of the instanton in the effective potential is just a displacement of the holomorphic function $E$. Therefore the dual geometry coincides with the analysis performed without instanton corrections. This is a common point with observations of the dualities in the $(2,2)$ GLSMs \cite{CaboBizet:2017fzc}.

\vskip 2truecm
\section{A model with global symmetry $SU(2)\times SU(2)$}
\label{sec5}

In this section we study a generalization of the model presented in \cite{Adams:2003zy} which consist of a GLSM with gauge symmetry $U(1)_1 \times U(1)_2$, two chiral fields $\Phi_1$, $\Phi_2$ and two Fermi $\Gamma_1$, $\Gamma_2$ with charge 1 under the first factor of the gauge symmetry $U(1)_1$; as well as two chiral fields $\widetilde\Phi_1$, $\widetilde\Phi_2$ and two Fermi $\widetilde\Gamma_1$, $\widetilde\Gamma_2$ with charge 1 under the $U(1)_2$ gauge group. This a deformation of a $(2,2)$ model into a $(0,2)$ model, so the restrictions for the fields $E$'s are:
\begin{eqnarray}
E_1=\sqrt2\{\Phi_1\Sigma+\widetilde\Sigma(\alpha_1\Phi_1+\alpha_2\Phi_2)\},\nn\\
E_2=\sqrt2\{\Phi_2\Sigma+\widetilde\Sigma(\alpha_1'\Phi_1+\alpha_2'\Phi_2)\},\nn\\
\widetilde E_1=\sqrt2\{\widetilde \Phi_1\widetilde \Sigma+\Sigma(\beta_1\widetilde \Phi_1+\beta_2\widetilde \Phi_2)\},\nn\\
\widetilde E_2=\sqrt2\{\widetilde \Phi_2\widetilde \Sigma+\Sigma(\beta_1'\widetilde \Phi_1+\beta_2'\widetilde \Phi_2)\},
\end{eqnarray}
where $\alpha$, $\alpha'$, $\beta$ and $\beta'$ are real parameters. In the limit when the $\alpha$'s and $\beta$'s parameters vanish the reduced $(0,2)$ model is recovered. The Lagrangian of this model is given by \cite{Adams:2003zy}
\begin{eqnarray}&&\boldsymbol{\Delta} L_{\tt original}=\sum_{i=1}^2\I\bigg\{-\frac i2\overline\Phi_i\bigg(e^{2\Psi}\partial_--\overleftarrow\partial_-e^{2\Psi}\bigg)\Phi_i+Ve^{2\Psi}|\Phi_i|^2-\frac12e^{2\Psi}\overline\Gamma_i\Gamma_i\bigg\}\nn\\
&+&\sum_{i=1}^2\I\bigg\{-\frac i2\overline{\widetilde\Phi}_i\bigg(e^{2\widetilde\Psi}\partial_--\overleftarrow\partial_-e^{2\widetilde\Psi}\bigg)\widetilde\Phi_i+Ve^{2\widetilde\Psi}|\widetilde\Phi_i|^2-\frac12e^{2\widetilde\Psi}\overline{\widetilde\Gamma}_i\widetilde\Gamma_i\bigg\}.
\end{eqnarray}

While the  potential of this Lagrangian is given by
\begin{equation}
U_{\tt original}=\frac{e^2}{2}\bigg(|\phi_1|^2+|\phi_2|^2-r_1\bigg)^2+\frac{e^2}{2}\bigg(|\widetilde\phi_1|^2+|\widetilde\phi_2|^2-r_2\bigg)^2+|E_1|^2+|E_2|^2+|\widetilde E_1|^2+|\widetilde E_2|^2.
\end{equation}
The vacuum solution for this model is \cite{Adams:2003zy}:
\begin{equation}
|\phi_1|^2+|\phi_2|^2=r_1,\qquad|\widetilde\phi_1|^2+|\widetilde\phi_2|^2=r_2,
\end{equation}
i.e., the vacua manifold is a product of $\mathbb P^1\times\mathbb P^1$ with Kähler classes $r_1$ and $r_2$ respectively, and
\begin{equation}
 E_i=\widetilde E_i=0.
\end{equation}
In the $SU(2) \times SU(2)$ generalization both chiral fields and Fermi fields are $SU(2)$ multiplets related a different $SU(2)$ sector.  Let us write the dual Lagrangian
\begin{eqnarray}
\boldsymbol{\Delta} L_{\tt master}&=&\I \sum_{i=1}^2 \frac{1}{8e_i^2}\overline\Upsilon_i\Upsilon_i+\int\text d\theta^+ \sum_{i=1}^2 \frac{t_i}{4}\Upsilon_i|_{\bt=0} \nn \\
&-&\I  \frac i2\overline\Phi e^{2 \Psi_1+2 \Psi_{1a}T_a}\bigg(\partial_{-}+iV_1+iV_{1a}T_a\bigg)\Phi  \nonumber \\
&-&\I  \frac i2\overline{\widetilde{\Phi}} e^{2 \Psi_2+2 \Psi_{2a}T_a}\bigg(\partial_{-}+iV_1+iV_{2a}T_a\bigg)\widetilde\Phi  \nonumber \\
&+&\I\bigg\{\frac i2\overline\Phi\bigg(\overleftarrow{\partial}_{-}-i V_1-i V_{1a}T_a\bigg)e^{2\Psi_1+2\Psi_{1a}T_a}\Phi\bigg\}  \nn \\
&+&\I\bigg\{\frac i2\overline{\widetilde{\Phi}}\bigg(\overleftarrow{\partial}_{-}-i V_2-i V_{2a}T_a\bigg)e^{2\Psi_2+2\Psi_{2a}T_a}\widetilde{\Phi}\bigg\}  \nn \\
&-& \I\bigg\{\  \frac{1}{2}\big(\overline \Gamma+\overline \Gamma_{1}\big)e^{2 \Psi_1+2 \Psi_{1a}T_a}\big(\Gamma+\Gamma_{1}\big)\bigg\} \nonumber  \\
&-& \I\bigg\{\  \frac{1}{2}\big(\overline {\widetilde\Gamma}+\overline {\widetilde\Gamma}_{2})e^{2 \Psi_2+2 \Psi_{2a}T_a}(\widetilde\Gamma+\widetilde\Gamma_{2})\bigg\} \nonumber  \\
&+& \sum_{i=1}^2 \I \Tr( \Lambda_i \Upsilon_i)+ \sum_{i=1}^2\I \Tr(\widetilde \Lambda_i \Upsilon_i) + {\rm h.c.}\nonumber\\
&+& \sum_{i=1}^2\I \overline{\chi}_i  E_i+ \sum_{i=1}^2\I \overline{\widetilde\chi}_i  E_i + {\rm h.c.}
\end{eqnarray}

Let us consider the following ansatz for the deformation of the (2,2) model in which $\alpha$ and $\beta$ are the parameters of the deformation:
{\small
\begin{eqnarray}
\left(\begin{array}{c}E_1 \\ E_2\end{array} \right)=\Sigma_0 \left(\begin{array}{c}\Phi_1 \\ \Phi_2\end{array} \right)+\widetilde\Sigma_0  \left(\begin{array}{c}\Phi_1 \\ \Phi_2\end{array} \right)\alpha_1+\Sigma  \left(\begin{array}{c}\Phi_1 \\ \Phi_2\end{array} \right)\alpha_2,\nn\\
\left(\begin{array}{c}\widetilde E_1 \\\widetilde E_2\end{array} \right)=\widetilde\Sigma_0 \left(\begin{array}{c}\widetilde\Phi_1 \\ \widetilde\Phi_2\end{array} \right)+\Sigma_0 \left(\begin{array}{c}\widetilde\Phi_1 \\ \widetilde\Phi_2\end{array} \right)\beta_1+\widetilde\Sigma\left(\begin{array}{c}\widetilde\Phi_1 \\ \widetilde\Phi_2\end{array} \right)\beta_2,\nn\\
\end{eqnarray}
This implies that $(E_1,E_2)$ and $(\widetilde E_1, \widetilde E_2)$ are vectors under $SU(2)_1$ and $SU(2)_2$ respectively. As well $(E_1,E_2)$  and $(\widetilde E_1, \widetilde E_2)$ are charged with charges 1 under the $U(1)_1$ and $U(1)_2$ respectively. Then, the dual Lagrangian becomes: 
\begin{eqnarray}
\boldsymbol{\Delta}L_{\tt dual}=\I\bigg\{&&\Big[Ve-Ve\frac{Z^aZ_a}{Z_0}-\frac{VY_+^aZ^a}{Z_0}\Big]+\BLF^\dagger \Big(e\mathbf{I_d}-\frac{\T^a}{Z_0}(eZ^a+2Y_+^a)\Big)^{-1}\BLF\nn\\
&+&\Big[\widetilde V\widetilde e-\widetilde V\widetilde e\frac{\widetilde Z^a\widetilde Z_a}{\widetilde Z_0}-\frac{\widetilde V\widetilde Y_+^a\widetilde Z^a}{\widetilde Z_0}\Big]+\BLF^\dagger \Big(\widetilde e\mathbf{I_d}-\frac{\T^a}{\widetilde Z_0}(\widetilde e\widetilde Z^a+2\widetilde Y_+^a)\Big)^{-1}\BLF \bigg\}\nn \\
&+&\frac{t}{4}\int\text d\theta^+\Upsilon|_{\bt=0}-\int\text d\theta^+\bigg[(\Sigma_0+\alpha_1\widetilde\Sigma_0)\BLF^\dagger\BLPhi+\alpha2\BLF^\dagger\Sigma\BLPhi\bigg] \nn\\
&+&\frac{\widetilde t}{4}\int\text d\theta^+\widetilde \Upsilon|_{\bt=0}-\int\text d\theta^+\bigg[(\widetilde \Sigma_0+\beta_1\Sigma_0)\BLF^\dagger\tilde \BLPhi+\beta2\BLF^\dagger\widetilde \Sigma\widetilde \BLPhi\bigg],
\end{eqnarray}
where the $\BLPhi$ is fixed (in terms of the $Z$-parameters) with (\ref{phis}).\\

For the scalar potential we have:
\begin{eqnarray}
U_{\tt dual}&=&-e\bigg(-y_+^aZ_a+\Im(t)\bigg)^2-\tilde e\bigg(-\tilde y_+^a\tilde Z_a+\Im(\tilde t)\bigg)^2 \nn\\
&+&\frac{1}{2y_+^ay_{a+}}\bigg[\overline H_1H_1+\overline H_2H_2+\bigg(\overline H_1H_1-\overline H_2H_2\bigg)\bigg(Z^3+2y_+^3\bigg)\nn\\
&&\qquad\qquad+\overline H_2H_1\bigg(2\overline{w}+\overline{Z}^{12}\bigg)+\text{h.c.}\bigg]\nn\\
&+&\frac{1}{2\widetilde y_+^a\widetilde y_{a+}}\bigg[\overline{\widetilde H}_1\widetilde H_1+\overline{\widetilde H}_2\widetilde H_2+\bigg(\overline{\widetilde H}_1\widetilde H_1-\overline{\widetilde H}_2\widetilde H_2\bigg)\bigg(\widetilde Z^3+2\widetilde y_+^3\bigg)\nn\\
&&\qquad\qquad+\overline{\widetilde H}_2\widetilde H_1\bigg(2\overline{w}+\overline{Z}^{12}\bigg)+\text{h.c.}\bigg]\nn\\
&+&\sqrt2\bigg[(\sigma_0+\alpha_1\widetilde\sigma_0)(\overline H_1\phi_1+\overline H_2\phi_2)+\alpha_2\overline H_1(\sigma^{11}\phi_1+\sigma^{12}\phi_2)\nn\\
&&\qquad\qquad+\alpha_2\overline H_2(\sigma^{21}\phi_1+\sigma^{22}\phi_2)+\text{h.c}\bigg]\nn\\
&+&\sqrt2\bigg[(\widetilde \sigma_0+\beta_1\sigma_0)\bigg(\overline{\widetilde H}_1\widetilde \phi_1+\overline{\widetilde H}_2\widetilde \phi_2\bigg)+\beta_2\overline{\widetilde H}_1\bigg(\widetilde \sigma^{11}\widetilde \phi_1+\widetilde \sigma^{12}\widetilde \phi_2\bigg)\nn\\
&&\qquad\qquad+\beta_2\overline{\widetilde H}_2\bigg(\widetilde \sigma^{21}\widetilde \phi_1+\widetilde \sigma^{22}\widetilde \phi_2\bigg)+\text{h.c}\bigg].
\end{eqnarray}
Thus, the bosonic scalar potential depends of 6 coordinates $y_+^a$ and $\widetilde y_+^a$, and the vacua $U_{\tt dual}=0$ after the minimum condition for $H$'s gives:
\begin{eqnarray}
U_{\tt dual}=\frac e2\bigg(\Im(t)-y_+^aZ_a\bigg)^2+(\overline E_1\quad\overline E_2)A\begin{pmatrix}E_1\\E_2\end{pmatrix} \nn \\
+\frac e2\bigg(\Im(\overline t)-\overline y_+^a\widetilde Z_a\bigg)^2+({\widetilde{\overline E }}_1\quad\widetilde {\overline E}_2)\overline A\begin{pmatrix}{\widetilde{\overline E }}_1\\{\widetilde{\overline E }}_2\end{pmatrix}=0,
\end{eqnarray}
with $A=\frac{-2y_+^by_+^b}{1-u_cu^c}\begin{pmatrix}u_3-1&u_{12}\\\overline u_{12}&-1-u_3\end{pmatrix}$,   $\widetilde A=\frac{-2\widetilde y_+^b\widetilde y_+^b}{1-\widetilde u_c\widetilde u^c}\begin{pmatrix}\widetilde u_3-1&\widetilde u_{12}\\ \widetilde {\overline u}_{12}&-1-\widetilde u_3\end{pmatrix}$ and
\begin{eqnarray}
E_1&=&\big[\sigma_0+\alpha_1\widetilde\sigma_0+\alpha_2(\sigma_{11}+\sigma_{12})\big]\phi_1,\, E_2=\big[\sigma_0+\alpha_1\widetilde\sigma_0+\alpha_2(\sigma_{21}+\sigma_{22})\big]\phi_2\;.\nn
 \\
\widetilde E_1&=&\big[\widetilde \sigma_0+\beta_1\sigma_0+\beta_2(\widetilde \sigma_{11}+\widetilde \sigma_{12})\big]\widetilde \phi_1, \, \widetilde E_2=\big[\widetilde \sigma_0+\beta_1\sigma_0+\beta_2(\widetilde \sigma_{21}+\widetilde \sigma_{22})\big]\widetilde \phi_2\;,
\end{eqnarray}
which as in the previous case, each one is positive quadratic form when $0<1-u^au_a$ and $0<1-\widetilde u^a\widetilde u_a$. If it is this case, the solution is
\begin{eqnarray}\label{vacuasu2xsu2}
r=y_+^aZ^a,\quad&\text{   }&\qquad\widetilde r=\widetilde y_+^a\widetilde Z^a, \nn\\
\sigma_0+\alpha_1\widetilde\sigma_0+\alpha_2(\sigma_{11}+\sigma_{12})=0,\quad&\text{   }&\quad \sigma_0+\alpha_1\widetilde\sigma_0+\alpha_2(\sigma_{21}+\sigma_{22})=0,\nn\\
\widetilde \sigma_0+\beta_1\sigma_0+\beta_2(\widetilde \sigma_{11}+\widetilde \sigma_{12})=0\quad&\text{and}&\quad \widetilde \sigma_0+\beta_1\sigma_0+\beta_2(\widetilde \sigma_{21}+\widetilde \sigma_{22})=0.
\end{eqnarray}
which is simply two copies of the potential with a single group factor.
For the instanton correction, it is
\begin{equation}
\int \text{d}\theta^+\bigg(\BLF^\dagger\boldsymbol{\beta} e^{\alpha^bY_b}+{\widetilde \BLF}^\dagger\widetilde{\boldsymbol{\beta}}e^{{\widetilde \alpha}^b\widetilde Y_b}\bigg).
\end{equation}
Thus the change for the scalar potential is given by: $\BLE\rightarrow \BLE-e^{\alpha_by_+^b}\boldsymbol{\beta}$. This means that the last 4 equations in (\ref{vacuasu2xsu2}) are equal to $|e^{\alpha_by_+^b}\boldsymbol{\beta}|^2$.

For the analyzed case the geometry of the dual model is the one of the product of two disks ${\bf D}_1 \times {\bf D}_2$,
which are the building blocks of the duality in subsection \ref{41}. Other
possible cases involve a not positive definite matrix $A$ or $\tilde{A}$. Notice also that the inclusion of instanton corrections preserves the geometry.

\vskip 1truecm
\section{Discussion and outlook}
\label{sec6}

In this work we describe T-dualities of two dimensional $(0,2)$ Abelian GLSMs. After a brief review on the basics of $(0,2)$ GLSMs, we started by constructing Abelian T-duality. This is implemented in models with $U(1)$ global symmetries; by gauging them. We analyse two cases: models coming from a $(2,2)$ supersymmetry reduction and pure $(0,2)$ models. The fundamental difference is that in the first case (reduction) the Fermi multiplet is dualized, while in the second case it is not. We study the simple example of a model with two chiral superfields, the first chiral field is charged under the $U(1)$ global symmetry and the other remains as an spectator, which just assists to obtain the  global symmetry. A master Lagrangian is obtained by promoting the global symmetry to be local, and adding Lagrange multiplier fields. The equations of motion for the gauge fields are obtained from the master Lagrangian leading to the dual action. The original chiral fields are eliminated by the gauge fixing procedure. We then compute the contributions to the scalar potential for all the terms in the dual Lagrangian. From the potential we determine the geometry of the space of supersymmetric  vacua. The geometry of the vacua space for the original model in both cases is $\mathbb{P}^1$. The dual model, under a single $U(1)$ T-duality,  has the topology of $\mathbb{R}^+\times \mathbb{R}$ for both cases. Notice that this is very different to the standard Mirror Symmetry duality, which will be a T-dualization of both chiral superfields. The instanton contributions to the superpotential are known for  $(0,2)$ models coming from a $(2,2)$ reduction \cite{Adams:2003zy,Melnikov2019}. For the case of a pure $(0,2)$ model, we still haven't obtained them, we leave this for future work. From our results it seems that there is a difference of considering $(0,2)$ models and their dual counterparts, if they come from a reduction or not. Moreover, in Section \ref{sec3} we carried out the duality algorithm for a model with two global Abelian symmetries \cite{Adams:2003zy}. This is a model which was later generalized in Section \ref{sec5} to the non-Abelian T-duality case. It consists of a reduction $(0,2)$ GLSM with gauge symmetry $U(1) \times U(1)$, six chiral superfields and six Fermi fields. The global Abelian symmetry is given by $U(1)^4$. 

Furthermore we construct T-dualities for Abelian GLSMs which non-Abelian global symmetry. Here we considered only the case coming from the reduction of a $(2,2)$ model. To be as general as possible, we obtain the master Lagrangian of a model with $U(1)^m$ gauge symmetry and  non-Abelian global group $G_1 \times \dots \times G_S$.  Starting from an original $(2,0)$ model with chiral, fermi superfields and a gauge multiplet, we obtain the master Lagrangian by gauging the global symmetries. We find suitable variables to write down the original master Lagrangians as a sum over the $S$ factors of the gauged symmetry group; then we find the equations of motion for the gauge fields. We considered a particular case with just one global $G=SU(2)$ factor and $U(1)$ gauge symmetry. The dual action is obtained by gauging the global symmetry $SU(2)$. It is observed that under a suitable redefinition of the chiral superfields in terms of new variables (fields) $Z$'s and $u^a$, the dual action can be rewritten in a simpler form.  In these variables also was found the scalar potential (\ref{potentiallast}). We also identify the conditions for which the potential is definite positive. This lead us to consider 3 regions depending on whether $y^a_+y^a_+$ is less than, equal or greater to $Z^a y^a_+$. We argued that these regions correspond to the condition with the topology of a open ball, a two-sphere or to the outside part of the sphere, respectively. Thus the vacua manifold for a positive semidefinite scalar potential corresponds to the closed disk ${\bf D}$. If the potential is not definite positive the component of the vacua manifold is $\mathbb{R}^2\diagdown\mathbf{D}$. Furthermore, we discussed  non-perturbative corrections to the superpotential via instantons. We find that if the instanton corrections are incorporated in the potential $U_{\tt dual}$ the effect is equivalent to shift $\BLE$ function as $\BLE - e^{\alpha_b y^b_+} \boldsymbol\beta$ in the potential without instanton corrections.
This coincides with the observation in the $(2,2)$ GLSMs non-Abelian T-duality were the instanton corrections preserve the dual geometry \cite{CaboBizet:2017fzc}.

In Section 5 we present the example of GLSM discussed in \cite{Adams:2003zy}, which comes
from a continuous $(0,2)$ deformation of a $(2,2)$ model. This model is a genuine pure $(0,2)$ GLSM. We worked out this model by gauging the global non-Abelian symmetry $SU(2)\times SU(2)$. We find the dual Lagrangian, and analyze the dual geometry of the vacua manifold. For the case of a positive definite potential the manifold is the Cartesian product of the vacua space of the $SU(2)$ simple model already discussed in Section \ref{sec4}, i.e. ${\bf D}_1 \times {\bf D}_2$. There are also instanton corrections affecting both sectors by a similar shifting of $\BLE$. 

To summarize we have constructed systematically non-Abelian T-duality in $(0,2)$ gauged linear
sigma models in two dimensions. In the future we would like to analyze more examples, given by
realistic Calabi-Yau manifolds, and their Mirror duals. We also are interested in analyzing models with a non-zero superpotential $J\not=0$, which will lead to compact Calabi-Yau. It would also be interesting to explore the connection of non-Abelian T-dualities in $(0,2)$ models with Mirror Symmetry in more general Calabi-Yau constructions (as Pfaffians and determinantal varieties).  We would like to explore the implications of these GLSMs non-Abelian T-dualities in string theory, as possible extensions of Mirror symmetry.

\newpage 

 \vspace{.5cm}
 \acknowledgments 
 

We thank Alejandro Cabo Montes de Oca, Stefan Groot Nibbelink, Kentaro Hori, Hans Jockers,  Leopoldo Pando Zayas, Martin Ro\v{c}ek,  Roberto Santos Silva, Erick Sharpe for useful comments and discussions. 

NGCB, JDC and HGC thank the support of the University of Guanajuato grant CIIC 264/2022 
"Geometría de dimensiones extras en teoría de cuerdas y sus aplicaciones físicas". NGCB would like to thank the Grant CONAHCyT A-1-S-37752, “Teorías efectivas de cuerdas y sus aplicaciones
a la física de partículas y cosmología” and  University of Guanajuato Grant CIIC
224/2023 "Conjeturas de gravedad cuántica y el paisaje de la teoría de cuerdas". NGCB would like to acknowledge 
support from the ICTP through the Associates Programme (2023-2029), and the Isaac Newton Institute for Mathematical Sciences, Cambridge, for support and hospitality during the programme "Black holes: bridges between number theory and holographic quantum information" where work on this paper was undertaken. This work was supported by EPSRC grant no EP/R014604/1. NGCB would like to thank the Simons Center for Geometry and Physics and the organizers of the conference "Gauged Linear Sigma Models @30" for the opportunity to present work related to this research.  JDC would
like to thank CONAHCyT for the support with a PhD fellowship.

\appendix

\section{Abelian T-duality algorithm in superfield components}
\label{appA}

In this appendix, we implement the Abelian dualization of a $(0,2)$ GLSM coming from a $(2,2)$ reduction in terms of superfield components. This is  as an alternative way to the superfield language, to carry out the duality. First, we write down the component expansion of the fields. The gauge superfield is given by
\begin{equation}
V=v_--2i\te\overline{\lambda}_--2i\bt\lambda_-+2\te\bt D,\qquad\Psi=v_+\te\bt.
\end{equation}
The fields in the model, including the chiral superfield, the Fermi superfield and the superfield $E$ are written as:
\begin{equation}
\widetilde\Phi=\phi+\sqrt2\theta^+\psi_+-i\theta^+\overline{\theta}^+(\partial_++iv_+)\phi,
\end{equation}
\begin{equation}
\Gamma=\gamma_--\sqrt2G\te-i\te\bt\partial_+\gamma_--\sqrt2E\bt,
\end{equation}
\begin{equation}
E(\Phi)=E(\phi)+\sqrt2\te\frac{\partial E}{\partial\phi}\psi_+-i\te\bt\partial_+E,
\end{equation}
The field component expansion for the Lagrangian multipliers which have been used in the bulk of the article are:
\begin{equation}
    \chi=x+\xi\te+\rho\bt+z\te\bt,
\end{equation}
\begin{equation}
    \Lambda=\omega+k\te+l\bt+\varepsilon\te\bt.
\end{equation}
The new dual fields are given by
\begin{equation}
Y_\pm=y_\pm+\sqrt2(\te\overline\xi_++\xi\bt)-i\te\bt\partial_+y_\mp,
\end{equation}
\begin{equation}
F=\eta_--\sqrt2\te  H_a-i\te\bt\partial_+\eta_-.
\end{equation}


Then the appropriate Lagrangian of a single chiral field and a Fermi one with Abelian global symmetry in component fields is written as
\begin{eqnarray}
L_{\tt components}&=&\overline\phi\phi D+i\overline\psi_+(\partial_-+iv_-)\psi_+-\sqrt2i(\lambda_-\psi_+\overline\phi-\overline\psi_+\overline\lambda_-\phi) \nn \\
&-&\frac12[\overline\phi(\partial_-+iv_-)(\partial_++iv_+)\phi-(\partial_++iv_+)\overline\phi(\partial_-+iv_-)\phi]+D\Im(t)+v_{01}\Re(t)\nn\\
&+&i\overline\gamma(\partial_++iv_+)\gamma+|G|^2-|E|^2-\left(\overline\gamma\frac{\partial E}{\partial\phi}\psi_++\frac{\partial \overline E}{\partial\overline\phi}\overline\psi_+\gamma\right)+\frac{v_{01}^2+D^2}{2e^2}.
\end{eqnarray}

To realize the T duality algorithm, gauging the global symmetry we add  the fields $v_\pm$, $\lambda_-$, $D$, $\gamma$ and $G$  (components of $V$, $\Psi$ and $\Gamma$) as well as the Lagrange multipliers. The original fields will be denoted with a subindex 0, the gauged ones with a subindex 1 and the sum of both without any subindex. Thus, for example: $a:=a_0+a_1$, etc. Besides it is convenient to define: $\delta_{\pm}:=\partial_{\pm}-\overleftarrow\partial_{\pm}$, thus:
\begin{eqnarray} \label{LagComp}
L_{\tt master}&=&\bar\phi\phi D+i\overline\psi_+(\partial_-+iv_-)\psi_+-\sqrt2i(\lambda_-\psi_+\overline\phi-\overline\psi_+\overline\lambda_-\phi) \nn\\
&-&\frac12\bigg[\overline\phi(\partial_-+iv_-)(\partial_++iv_+)\phi-(\partial_++iv_+)\overline\phi(\partial_-+iv_-)\phi\bigg]+D\Im(t)+v_{01}\Re(t)\nn\\
&+&i\overline\gamma(\partial_++iv_+)\gamma+|G|^2-|E|^2-\left(\overline\gamma\frac{\partial E}{\partial\phi}\psi_++\frac{\partial \overline E}{\partial\overline\phi}\overline\psi_+\gamma\right)+\frac{v_{01}^2+D^2}{2e^2}\nn\\
&+&2iD(\omega-\overline\omega)+v_{01}(l+\overline{} l)+2(\varepsilon\lambda_-+\overline\lambda_-\overline\varepsilon)-2i(l\partial_+\lambda_-+\partial_+\overline\lambda_-\overline l)\nn\\
&+&i(\overline x\partial_+E-x\partial_+\overline E)-\sqrt2\bigg(\overline\xi\frac{\partial E}{\partial\phi}\psi_++\overline\psi_+\overline{\frac{\partial E}{\partial\phi}}\xi\bigg)-\overline zE-z\overline E.
\end{eqnarray}
Taking variations with respect to $v_{\pm,1}$, $\lambda_1$, $D_1$, $\gamma_1$ and $G_1$ one obtains the corresponding equations of motion:


For $\delta_{D_1}L$:
\begin{equation}\label{varD}
i(l-\overline l)=2|\phi|^2.
\end{equation}

For $\delta_{\lambda_1}L$:
\begin{equation}\label{varl}
\varepsilon+2i\partial_+\omega=4\overline\phi\psi_+.
\end{equation}

For $\delta_{b_1}L$:
\begin{equation}\label{varb}
4v_+|\phi|^2=4\overline\psi_+\psi_++2i(\overline\phi\delta_+\phi)+\partial_{++}(l+\overline l).
\end{equation}

For $\delta_{a_1}L$:
\begin{equation}\label{vara}
4v_-|\phi|^2=-\frac i2\overline\gamma\gamma+2i(\overline\phi\delta_-\phi)-\partial_{--}(l+\overline l)+\overline\gamma\gamma .
\end{equation}

For  $\delta_{\gamma_1}L$:
\begin{equation}\label{varg}
(\partial_+-iv_++2v_+)(\overline\gamma_0+\overline\gamma_1)=i(\mu\overline\phi+\overline s\overline\psi_+).
\end{equation}

For $\delta_{G_1}L$:
\begin{equation}\label{varG}
G_0+G_1=0.
\end{equation}
New variables can be defined in the form:
\begin{equation}
l_\pm:=l\pm\overline l,\qquad y:=-i\varepsilon+2\partial_{++}\omega\qquad\text{and}\qquad f:=z+2i\partial_{++}x\;.
\end{equation}
Thus, using Eqs. (\ref{varD}, \ref{varl}, \ref{varb}, \ref{vara}) in the Lagrangian (\ref{LagComp}) it results the dual Lagrangian:
$$
\frac12L_{\tt dual} \ \ = \ \ -i(y\lambda_{0-}-\overline\lambda_{0-}\overline y)+\frac12(fs_0+\overline f\overline s_0)-iD_0l_--b_0\partial_+l_++a_0\partial_-l_+
$$
\begin{equation}
=-i(\mu_0\overline\rho-\rho\overline\mu_0)+\overline\phi\delta_-\delta_+\phi+4ab|\phi|^2-|G|^2-2i\overline\psi_+\delta_-\psi_+-\frac i2\overline\gamma\delta_+\gamma.
\end{equation}
It is useful to emphasize that $a=a_0+a_1$ and $b=b_0+b_1$. It is easy to check that this dual Lagrangian coincides with the component field expansion of the dual Lagrangian (\ref{dual}). A similar procedure could be carried over in the case of models with non-Abelian T-duality.
\newpage

\bibliography{biblioNatd}

\end{document}